\documentclass{article}
\usepackage{graphicx}
\usepackage{amsmath}
\usepackage{amssymb}
\usepackage{multicol}
\usepackage{etoolbox}
\usepackage{url}
\usepackage{float}
\usepackage{booktabs}

\usepackage[left=0.9in,right=0.9in,top=0.9in,bottom=0.9in]{geometry}

\usepackage[numbers,sort&compress]{natbib}

\usepackage{setspace}
\usepackage{authblk}

\usepackage[switch,modulo]{lineno}

\usepackage{xr}
\title{Mobility shapes heat exposure inequalities in cities}    
\author[1,*]{Marc Duran-Sala}
\author[2]{Mattia Mazzoli}
\author[1]{Martin Hendrick}
\author[1,*]{Gabriele Manoli}
\affil[1]{Laboratory of Urban and Environmental Systems, École Polytechnique Fédérale de Lausanne, Switzerland}
\affil[2]{ISI Foundation, Turin, Italy}
\affil[*]{Corresponding author(s): marc.duransala@epfl.ch, gabriele.manoli@epfl.ch}

\date{\today}

\begin{document}


\maketitle

\begin{abstract}

Segregation has long been recognized as a driver of environmental inequalities, with disadvantaged groups often living in neighborhoods where heat-related risks are highest. Yet, it remains unclear how daily mobility patterns, embedded within heterogeneous urban heat fields, shape heat exposure inequalities across sociodemographic groups. Using mobile phone records of daily mobility flows and urban temperature fields across 23 Spanish cities, we develop a network-based framework to quantify how different sociodemographic groups experience heat through their daily movements. We further apply the framework to tract-level commuting networks across 30 major US cities as an external validation, yielding qualitatively comparable patterns. We find systematic income-related inequalities, with low-income groups consistently experiencing higher exposure than high-income groups, while age-related disparities are smaller in magnitude. These inequalities intensify during commuting trips, indicating that routine mobility amplifies spatial heat gradients more than non-routine movements. 
Finally, we show that parsimonious population-based mobility models with group-agnostic mobility rules reproduce an important component of the observed exposure disparities, suggesting that these inequalities emerge from the interplay between the unequal spatial organization of daily activities across sociodemographic groups and urban heat gradients. Our findings provide a generalizable framework to characterize inequalities in mobility-based heat exposure across cities and inform climate-resilient urban planning and public health strategies under intensifying climate-related risks.


\end{abstract}

\section*{Introduction}

Built environments increase local temperatures and amplify the risks associated with extreme heat events \cite{Vicedo-Cabrera2021}, posing serious economic and public health challenges for cities. For example, high heat exposure is associated with increased risk of hospital admissions \cite{Zhang2015} and temperature-related mortality \cite{Huang2023} but also with reduced labor productivity \cite{Dasgupta2024, Burke2015} and diminished learning outcomes \cite{Learning2020}.
%
These impacts are unevenly distributed across the population as they depend both on the local climatic conditions \cite{Manoli2019} and the exposure and vulnerability of different sociodemographic groups, with older adults and individuals with pre-existing conditions facing higher mortality risks \cite[e.g.,][]{Kenny2010, Zanobetti2012, Zhang2015}.

At the same time, temperature fields vary within cities due to differences in the built-environment, such as the density of buildings, roads, and green spaces \cite[e.g.,][]{Oke2017,Li2024,Duran-Sala2026}. Urban segregation patterns drive inequality in environmental exposure, with specific racial or low-income neighbourhoods often experiencing higher heat levels \cite{Hsu2021, Chakraborty2019, Chakraborty2023}. While urban heat risk assessments have received increasing attention in recent years \cite{Dong2024}, most studies of urban heat exposure inequalities rely on static residence-based approaches \cite[e.g.,][]{Hsu2021, Chakraborty2019} that overlook the dynamic nature of exposure. Heat exposure depends not only on where people live, but also on where they spend their time throughout the day -- commuting, working, shopping, or engaging in leisure activities. Mobility patterns therefore play a central role in shaping how individuals interact with heterogeneous urban thermal environments. 

Recent studies incorporating mobility information have shown that daily movements can substantially reshape heat exposure patterns \cite[e.g.,][]{Lin2025}. For example, some metropolitan areas exhibit “heat traps”, where residents of high-heat neighborhoods predominantly visit similarly hot areas, while others display “heat escapes”, where mobility connects thermally dissimilar districts \cite{Huang2024}. However, how these mobility--heat dynamics translate into social inequalities remains less understood. City-specific studies incorporating mobility data report divergent conclusions on whether daily movement attenuates or amplifies exposure disparities. In some contexts, mobility-based exposure has a modest impact on residence-based exposure estimates \cite[e.g.,][]{Garber2025}, or reduces extremes relative to residence-based estimates, as mobility induces an averaging effect that moderates extreme residential conditions \cite[e.g.,][]{Wu2025}. In contrast, a recent study found that incorporating workplace exposure amplifies socioeconomic disparities \cite{Huang2025}.

Despite these advances, important gaps remain in understanding how daily mobility reshapes heat exposure inequalities across sociodemographic groups and urban contexts. Existing studies have typically examined mobility-based heat exposure either across cities but without sociodemographic group information, or with sociodemographic group information but within individual metropolitan areas. This makes it difficult to assess whether inequalities in mobility-based heat exposure follow consistent patterns across urban contexts under a common framework. This is important because incorporating attributes such as income, age, gender, or ethnicity into mobility analyses can alter estimates of spatial and social segregation \cite{Barbosa2021, Xu2025, Lenormand2015, duransala2024}. As a result, it remains unclear whether mobility systematically amplifies, attenuates, or preserves heat exposure inequalities across cities and sociodemographic groups.

%
Here, we address this knowledge gap by using mobile phone records describing daily origin–destination flows between districts stratified by income and age group, and high-resolution temperature fields for 23 Spanish cities during July-August 2022 and 2023 (see Fig.~\ref{fig:1}a and Supporting Information, SI Table~\ref{tab:temp_cities}). Specifically, we develop a network-based framework in which districts are ranked by temperature deciles and origin–destination flows define directed weighted mobility networks across urban heat fields. By focusing on relative thermal rankings, we characterize how sociodemographic groups are structurally positioned within each city’s internal heat hierarchy while enabling comparisons across cities with different climatic baselines. This framework allows us to quantify not only average heat exposure, but also the assortative and directional structure of mobility networks across the urban heat field, revealing how different population groups experience heat through their daily movements. To evaluate the generality of the framework and observed patterns, we further perform an independent validation using commuting networks across 30 major US cities (SI Table~\ref{tab:SI_US_cities_summary} and SI Fig.~\ref{fig:SI_illustrative_US}).

Moreover, to assess how much of the observed exposure inequalities can be reproduced using parsimonious trip allocation mechanisms, we implement two population-based mobility models: the gravity and radiation models \cite{Zipf1946, Simini2012}. Importantly, both models allocate inter-district flows using group-agnostic mobility rules based on population distributions and geographic distance. 

Overall, this study provides a generalizable framework to quantify how daily mobility redistributes exposure across urban heat fields and contributes to inequalities across sociodemographic groups. In the context of intensifying climate risks, such an approach can help inform more equitable climate-sensitive urban planning and public health strategies.

\section*{Results}

\begin{figure}[H]   
    \centering
    \includegraphics[width=\linewidth]{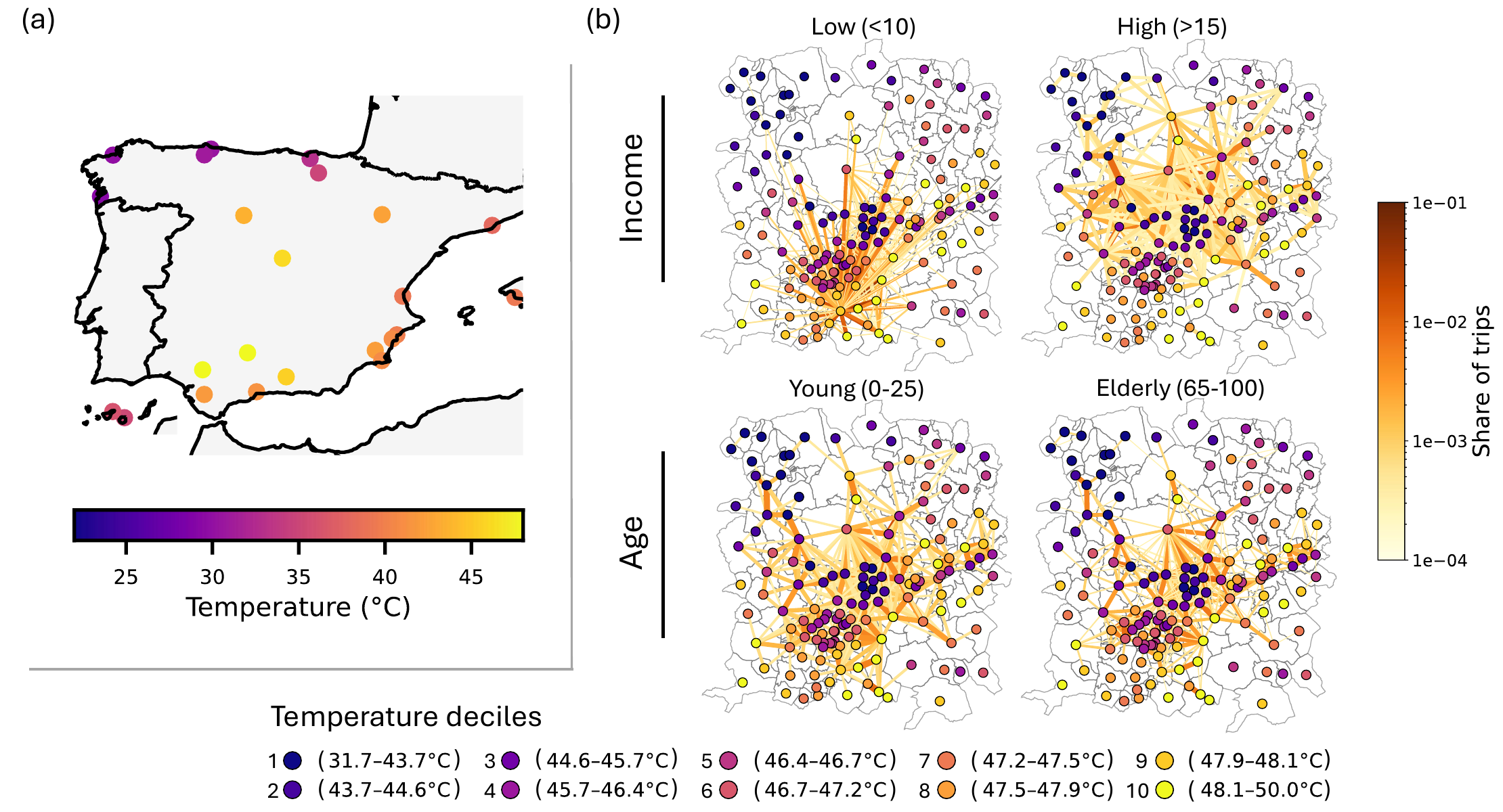}
    \caption{\footnotesize 
    \textbf{Mobility networks at the district level stratified by age and income groups across Spanish cities.}  (a) City-level mean temperature averaged over July–August 2022 and 2023 for the 23 cities included in the analysis. (b) Mobility network at district-level in Madrid aggregated over July–August 2022 and 2023. Only the top $5\%$ of links by share of trips are displayed, with edge width proportional to the absolute number of trips and edge color indicating the relative share of trips. Nodes are colored by deciles of average district temperature over the same period.}
\label{fig:1}
\end{figure}

We analyse daily mobility flows across 23 Spanish cities (Fig.~\ref{fig:1}a; SI Table~\ref{tab:SI_cities_summary}) using anonymized mobile phone data stratified by age and income, combined with high-resolution temperature fields at the district level during July-August 2022 and 2023 (see Methods for details, including the definition of city boundaries). Our focus is on daily experienced heat exposure and its intra-urban variations due to daily trips and activities, as compared to traditional heat exposure assessments based on residential environments only. Thus, we consider trips between districts and within the same district, but exclude those trips whose destination corresponds to a home-labeled activity. Including them would overweight the home district in our mobility-based metrics blurring the signal associated with exposure in workplaces, commercial areas, and other activity locations (see Methods).

To study how different age and income groups move across the urban temperature field, we build a network for each city and sociodemographic group, where districts are represented as nodes and weighted directed edges represent the number of trips between them (e.g., Fig.~\ref{fig:1}b). Each district is classified into a temperature decile based on its mean summer land surface temperature (July-August 2022 and 2023), providing a stable thermal ranking that preserves intra-urban contrasts while enabling consistent comparison of inequality patterns across cities (see Methods). Using these temperature deciles, we construct origin--destination (OD) matrices whose entries \(X_{ij}\) represent the aggregated number of trips from districts in temperature decile \(i\) to those in decile \(j\) (Fig.~\ref{fig:2}a-b). From these matrices we derive a set of mobility-based metrics capturing complementary aspects of how groups interact with the urban temperature field within each city: heat exposure $E$, which quantifies the mean temperature decile of visited destinations; heat assortativity $\rho$, which measures the tendency for trips to connect districts with similar temperature levels; and heat directionality $R$, which characterizes systematic biases in movement toward either hotter or cooler districts. To quantify exposure inequality across population groups, we compute the heat exposure difference between low- and high-income groups ($\Delta E^{inc}$) and between young and elderly age groups ($\Delta E^{age}$). For further details and formal definitions, the reader can refer to the Methods section.

\subsection*{Mobility patterns reveal systematic heat‐exposure inequalities across cities}

\begin{figure}[H]   
    \centering
    \includegraphics[width=\linewidth]{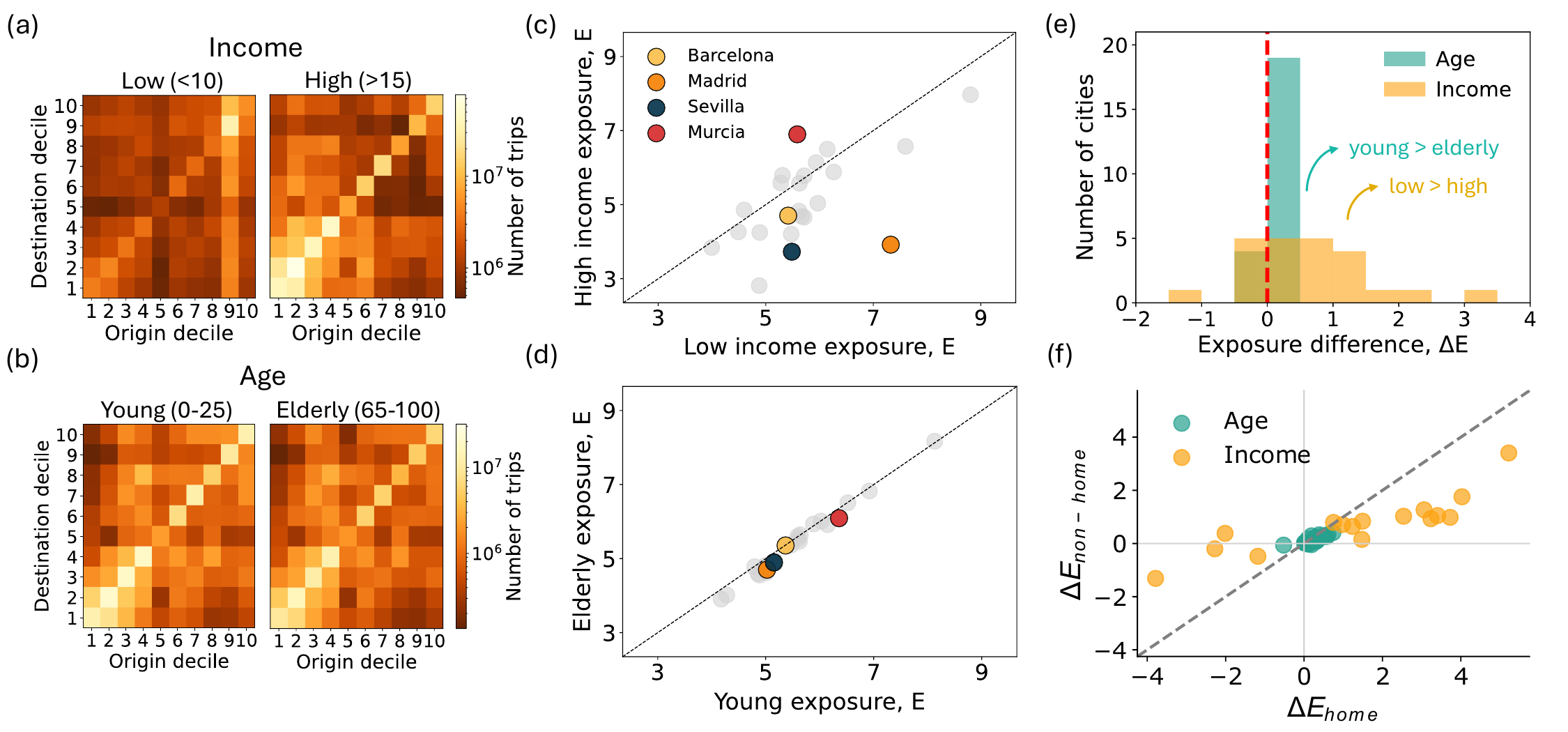} 
    \caption{\footnotesize 
    \textbf{Inequalities in mobility-based heat exposure across income and age groups.}
    (a-b) Origin-destination matrices (ODs) between temperature deciles for income and age groups in Madrid.
    (c-d) Experienced heat exposure of high versus low-income groups, and elderly versus young populations across cities. Colored points highlight selected cities for visual reference; grey points correspond to the remaining cities. The $y=x$ line indicates no exposure inequality between groups.
    (e) Histogram of exposure difference $\Delta E$ across cities. Positive values indicate that low-income groups and young populations are more exposed than high-income groups and elderly populations, respectively.
    (f) Home versus non-home exposure inequalities. Each point represents one city. The x-axis shows the exposure difference computed for trips ending at home ($\Delta E_{home}$), while the y-axis shows the corresponding difference for trips ending at non-home activities ($\Delta E_{non-home}$). Income contrasts are defined as low minus high income exposure, and age contrasts as young versus elderly population exposure. The dashed line indicates equal exposure inequality at home and non-home destinations; points below the line indicate smaller inequalities for non-home exposure. See SI Table~\ref{tab:SI_paired_joint_combined} for paired statistical tests across cities.}
\label{fig:2}
\end{figure}

Aggregating mobility flows over the study period, Fig.~\ref{fig:1}b exemplifies how mobility patterns of different population groups are distributed unevenly across the urban temperature field in Madrid (see SI for additional cities). Specifically, low-income groups tend to concentrate a larger share of their trips in high-temperature deciles, whereas high-income groups predominantly travel within lower-temperature deciles (Fig.~\ref{fig:2}a). Similar contrasts in the distribution of trips across temperature deciles are observed in most cities. Moreover, all cities exhibit a positive degree of heat assortative mobility across all income and age groups, with most trips occurring among districts of similar temperature deciles (see SI Table~\ref{tab:SI_city_metrics_income_age}). Such mobility patterns likely contribute to reinforcing pre-existing residential disparities: median neighborhood income decile is negatively correlated with local temperature decile in 19 out of the 23 cities (see SI Table~\ref{tab:structural_income_heat}), indicating that low-income areas tend to be systematically hotter (consistently with previous findings \cite[e.g.,][]{Hsu2021, Chakraborty2019}). Combining these residential conditions with mobility flows, we compute the mobility-based heat exposure $E$ (see Methods) across sociodemographic groups.

We find persistent income inequalities across most cities, with low-income groups experiencing higher heat exposure than high-income groups (two-sided Wilcoxon signed-rank test: $p=1.23 \times 10^{-2}$; SI Table~\ref{tab:SI_paired_joint_combined}). These differences remain significant within every age group, with the strongest differences observed among elderly populations (two-sided Wilcoxon signed-rank test: $p=4.85 \times 10^{-3}$; SI Table~\ref{tab:SI_paired_joint_combined}). This pattern is shown in Fig.~\ref{fig:2}c, where each point represents one city. Colored points highlight selected cities for visual reference; grey points correspond to the remaining cities. Inequality is measured by deviations from the $y=x$ dashed line: points below the diagonal indicate higher exposure for low-income people, while points above the line indicate higher exposure for high-income people in those cities. Madrid exhibits the largest exposure difference, with low-income group experiencing substantially higher exposure than high-income group, whereas Murcia shows the strongest reversed pattern, where the high-income group experiences higher exposure than the low-income group.

Age-related inequalities are remarkably consistent across cities, with younger groups exposed to slightly hotter conditions than elderly groups (two-sided Wilcoxon signed-rank test: $p=8.85 \times 10^{-5}$; SI Table~\ref{tab:SI_paired_joint_combined}). This pattern is visible in Fig.~\ref{fig:2}d by the systematic displacement of cities below the $y=x$ dashed line. However, these age-related inequalities are smaller than income-related disparities and become less consistent after stratifying by income group -- with no evidence of a systematic exposure difference between young and elderly populations among low-income groups (two-sided Wilcoxon signed-rank test: $p=0.61$; SI Table~\ref{tab:SI_paired_joint_combined}) -- indicating that income constitutes the dominant and most consistent axis of inequality in mobility-based heat exposure across cities.

The distribution of exposure differences across cities further highlights the systematic nature of these inequalities (Fig.~\ref{fig:2}e). For income, the distribution is skewed toward positive values $\Delta E^{inc}>0$, with several cities exhibiting substantial heat exposure disparities. In contrast, age-related exposure differences are more concentrated around zero and display smaller magnitudes (Fig.~\ref{fig:2}e).
Full city-level values for heat exposure, assortativity, and directionality are reported in SI Table~\ref{tab:SI_city_metrics_income_age}.

Placebo tests, where income and age labels are shuffled within each city-day while preserving the observed mobility structure, eliminate the exposure differences. For each permutation, we recomputed the city-level exposure differences and summarized them using the mean exposure difference across cities. Across $B=100$ permutations, the null distributions of the mean exposure difference were tightly centered around zero for both income and age contrasts, indicating that the observed inequalities are unlikely to arise spuriously from random associations between sociodemographic labels, mobility flows, and urban heat fields (empirical two-sided permutation test, $p=0.01$; SI Fig.~\ref{fig:SI_permutation_test}). We further tested robustness to noisy OD links by applying increasingly strict minimum-trip thresholds to remove low-volume flows, and found negligible impact on the magnitude or direction of the exposure differences (SI Fig.~\ref{fig:SI_thresholding_test}), indicating that the results are not driven by sparse mobility connections. To assess robustness to sampling variability in mobility flows, we performed a multinomial bootstrap on the aggregated OD trip counts within each city and sociodemographic group. Across all cities, the resulting confidence intervals remained narrow relative to the observed exposure differences (SI Table~\ref{tab:SI_bootstrap}).
Moreover, because the analysis relies on 23 city-level observations, we tested whether the results were sensitive to individual cities using leave-one-city-out analyses. Recomputing the average exposure differences after iteratively removing each city preserved both the sign and magnitude of the inequalities in all cases. Average income-related exposure differences ($\Delta E^{inc} \in [0.46,0.65]$) and age-related exposure differences ($\Delta E^{age} \in [0.14,0.16]$) remained positive across all leave-one-out realizations, indicating that the results are not sensitive to outliers.

To compare mobility-based exposure with residence-based approaches, we separately evaluated exposure from trips whose destination corresponds to a non-home-labeled activity versus exposure from trips whose destination corresponds to a home-labeled activity (Fig.~\ref{fig:2}f). Inequality magnitudes were larger for home-labeled destinations in nearly all cities, computed as $|\Delta E_{home}|-|\Delta E_{non-home}|$ (paired two-sided Wilcoxon signed-rank test: $p=6.10\times10^{-5}$ for income and $p=2.62\times10^{-4}$ for age). However, the sign of the inequality was preserved between home and non-home exposure, meaning cities where one group is more exposed at home destinations generally show the same group as more exposed at non-home destinations (sign test: $p=2.59\times10^{-4}$ for income and $p=2.44\times10^{-4}$ for age). These results suggest that heat-exposure inequalities are related to residential structure, while also revealing that these disparities extend across urban heat gradients through daily mobility.

Finally, as an external validation of the framework and to assess whether these aggregate inequality patterns generalize beyond the Spanish cities, we repeated the analysis for 30 major US cities (see SI Table~\ref{tab:SI_US_cities_summary} and SI Fig.~\ref{fig:SI_illustrative_US}), using tract-level home--work OD matrices and July--August 2022 daytime air-temperature data (see Methods). The results are qualitatively consistent, with low-income groups more exposed than high-income groups (SI Fig.~\ref{fig:SI_figure2_US}a; two-sided Wilcoxon signed-rank test: $p = 1.13 \times 10^{-2}$; SI Table~\ref{tab:SI_US_paired_tests}), while age-related exposure differences remain smaller in magnitude (SI Fig.~\ref{fig:SI_figure2_US}b; two-sided Wilcoxon signed-rank test: $p = 6.12 \times 10^{-2}$; SI Table~\ref{tab:SI_US_paired_tests}).

\subsection*{Commuting and non-routine mobility exhibit distinct daily structural dynamics}

\begin{figure}[H]   
    \centering
    \includegraphics[width=\linewidth]{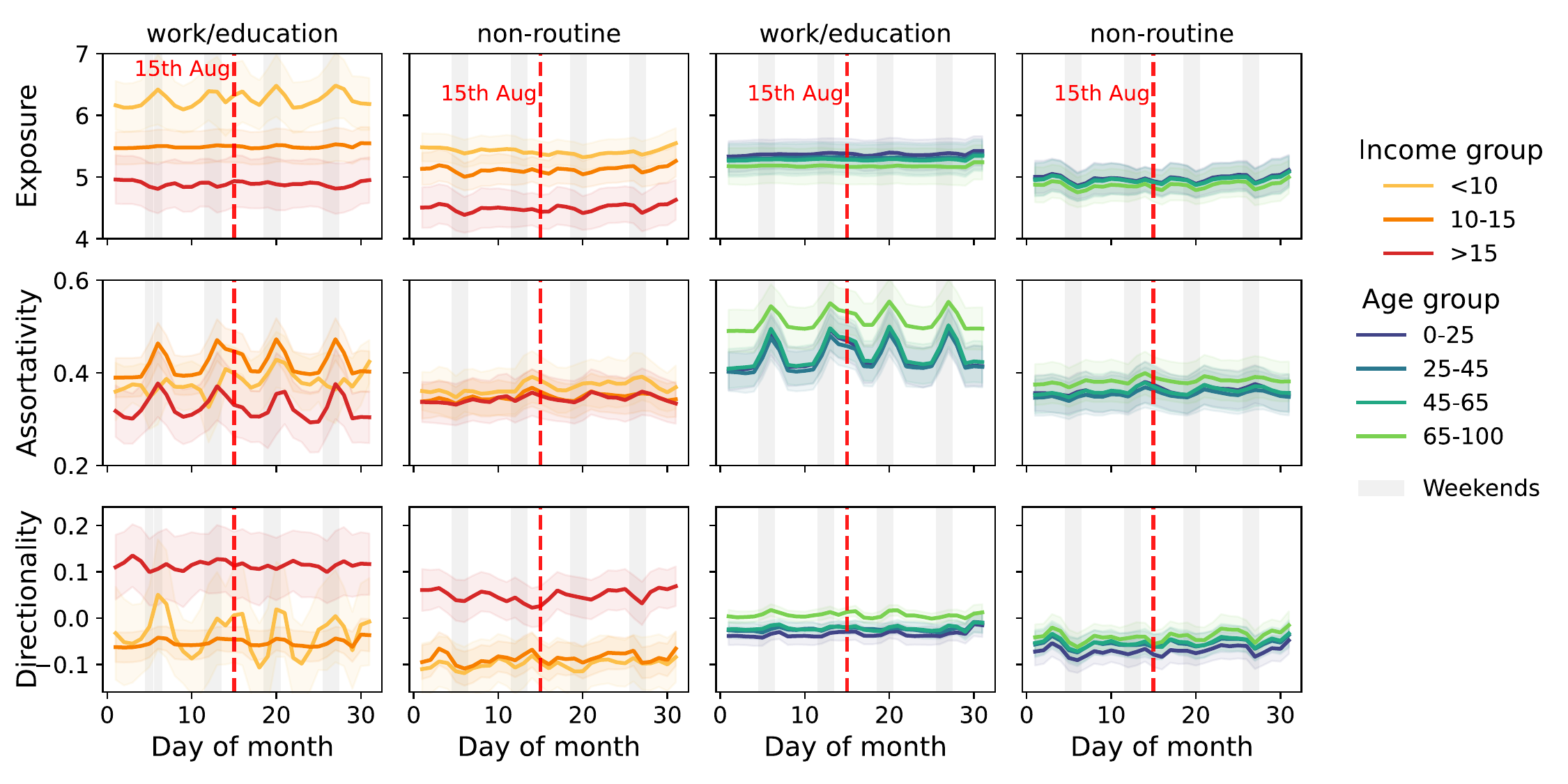} 
    \caption{\footnotesize 
    \textbf{Daily dynamics of mobility-based heat exposure, heat assortativity, and heat directionality differ between commuting and non-routine trips across income and age groups.}
    Daily population-weighted average heat exposure, assortativity, and directionality across all cities in August 2023, shown separately for work/education trips and non-routine trips, for income and age groups. Error bands represent the standard error of the population-weighted mean. Grey shading indicates weekends. The dashed red line marks 15 August, a Spanish national holiday. A 2-day centered window smooths short-term noise.}
\label{fig:3}
\end{figure}

Next we analyze mobility flows at the daily scale to assess the temporal persistence of these heat inequalities beyond summer averages. Using daily mobility records, we show that heat exposure, assortativity, and directionality exhibit stable temporal patterns, and that experienced heat inequalities remain persistent (Fig.~\ref{fig:3}; see SI Fig.~\ref{fig:SI_commute_vs_leisure} for other months).
Figure~\ref{fig:3} reports the daily population-weighted average of each metric across all cities, with each city contributing proportionally to its population size, so that the resulting average reflects the demographic relevance of each city (see Methods). Grey shaded areas indicate weekends, and the red dashed line marks 15 August, a national public holiday in Spain. Across income groups, exposure levels remain clearly separated on every day of the month, with the low-income group consistently visiting hotter districts than the high-income group. Age-related differences are smaller in magnitude but similarly stable over time. 

Importantly, these daily dynamics vary systematically by trip purpose (see Methods). Trips involving work or study activities (referred to as commuting trips) exhibit consistently higher heat exposure levels and more differentiated heat assortativity than trips involving infrequent activities (referred to as non-routine trips) across all days (Fig.~\ref{fig:3}). The exposure difference between low- and high-income groups ($\Delta E$) is also larger for commuting trips, indicating that routine mobility reinforces existing spatial heat gradients more strongly than non-routine mobility.
All three metrics display pronounced weekday–weekend cycles. Exposure declines slightly during weekends for non-routine trips, and increases during weekends for trips involving work or study activities among the low-income group. Heat assortativity increases during weekends for commuting trips, reflecting a shift from structured weekday commuting — which connects a broader range of temperature deciles — to more spatially homogeneous flows. Directionality shows a complementary pattern: low- and mid-income groups tend to move toward cooler districts, whereas high-income group more frequently travel toward hotter ones.
A marked structural perturbation is also visible around 15 August. On this day, mobility patterns partially resemble weekend dynamics. However, the relative ranking of heat exposure across income and age groups remains unchanged. 

\subsection*{Parsimonious mobility models reproduce aggregate heat-exposure inequalities}

To evaluate whether heat exposure inequalities observed in mobility flows aggregated over the full study period can emerge from parsimonious trip allocation mechanisms based on group-agnostic mobility rules, we construct synthetic OD flows using the gravity model \cite{Zipf1946} and the parameter-free radiation model \cite{Simini2012} (see Methods). These models provide a counterfactual test of how much of the observed exposure inequalities can be reproduced when destinations are allocated from population distributions and geographic distance, while preserving the observed group-specific total outflows. Comparing modelled and empirical exposure inequalities therefore allows us to assess whether observed disparities can already be approximated from group-specific outflows and generic mobility rules, without explicitly modelling detailed group-specific destination preferences.

\begin{figure}[H]   
    \centering
    \includegraphics[width=\linewidth]{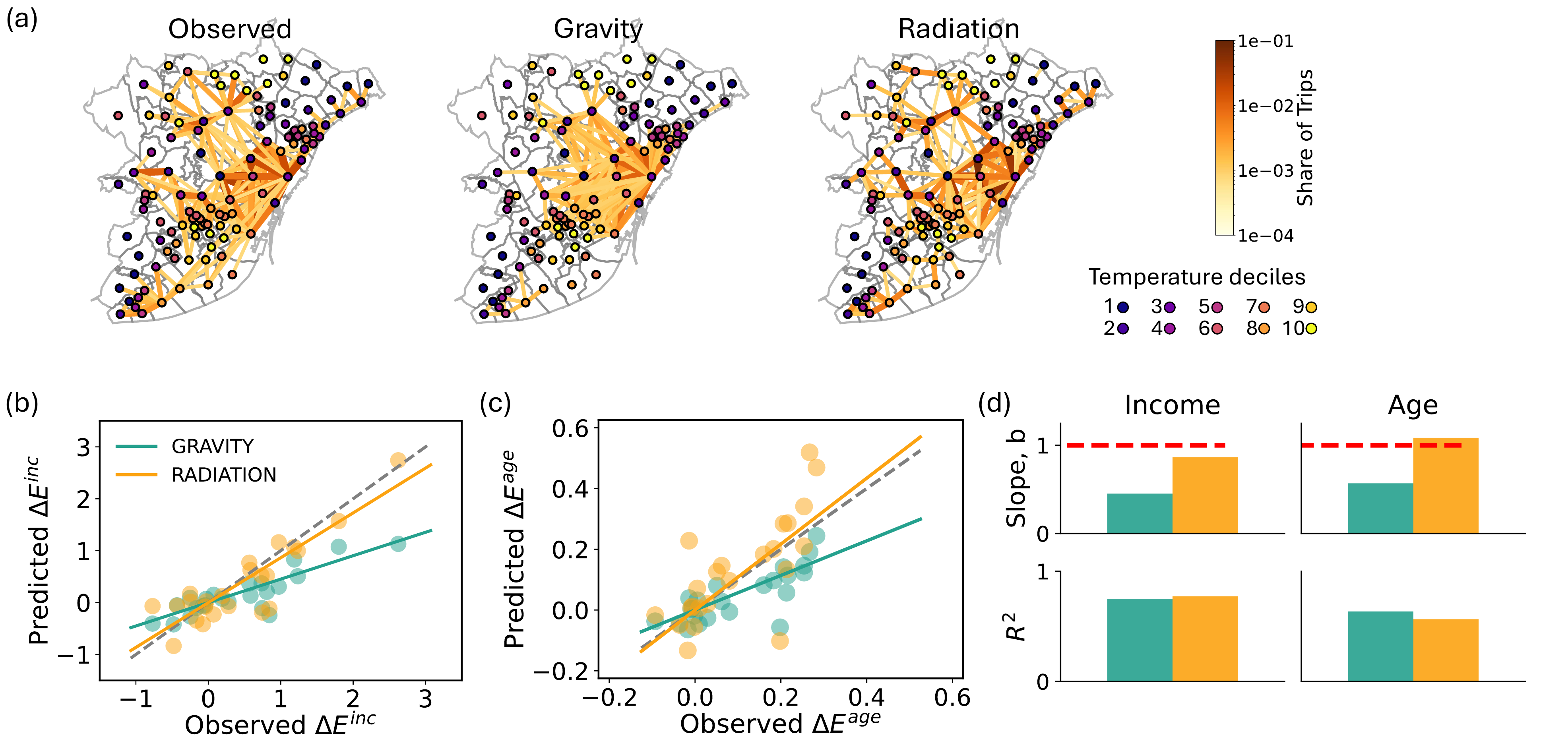} 
    \caption{\footnotesize \textbf{Parsimonious mobility models largely reproduce income- and age-related heat exposure inequalities.}
    (a) Illustration of the mobility networks derived from empirical data and from gravity and radiation models for Barcelona city during July-August 2022 and 2023 for high income group. Only the top $200$ links ranked by share of trips are displayed. Edge width scales with the absolute number of trips, while edge color encodes the relative share of trips. Nodes are colored by deciles of average district temperature over the same period.
    Observed versus predicted heat exposure differences for (b) income ($\Delta E^{inc}$) and (c) age ($\Delta E^{age}$) for each city. The dashed gray $y=x$ line indicates perfect agreement between observations and model predictions. (d) Summary of regression metrics from the linear fit of predicted vs observed exposure differences across cities. Bars show slope (top) and coefficient of determination $R^2$ (bottom).}
\label{fig:4}
\end{figure}

Because the modelled OD matrices exclude intra-district flows, empirical OD matrices were recomputed accordingly for all empirical-versus-modelled comparisons. For each city, we computed the income and age exposure differences ($\Delta E^{inc}$ and $\Delta E^{age}$) from both observed and modelled OD matrices, and regressed the modelled values against the observed values using a linear model (see Methods for details). The slope of this regression quantifies how much of the inequality magnitude is reproduced across cities (with $b=1$ indicating modelled and observed inequalities increase proportionally), while the intercept and bias capture systematic over- or under-estimation of the observed exposure difference (Fig.~\ref{fig:4}; regression metrics in SI Table~\ref{tab:SI_gap_regression_summary}).

Income-related exposure disparities span a wide range across cities (Fig.~\ref{fig:2}e). Both models recover a substantial fraction of the observed inequalities but their accuracy differs (see Fig.~\ref{fig:4}b). While the gravity model yields lower mean absolute errors in group-specific exposure levels (SI Table~\ref{tab:SI_E_regression_summary}), the radiation model better preserves the relative exposure contrasts between groups (SI Table~\ref{tab:SI_gap_regression_summary}). The gravity model underestimates the income exposure difference (Fig.~\ref{fig:4}d; SI Table~\ref{tab:SI_gap_regression_summary}), reflecting the exposure underprediction for the low-income group and slight overprediction for the high-income group in several cities (e.g., Madrid, Sevilla and Vigo; SI Fig.~\ref{fig:SI_E_income_per_city}), thereby compressing the observed exposure differences. In contrast, the radiation model reproduces the income difference much more closely (Fig.~\ref{fig:4}d; SI Table~\ref{tab:SI_gap_regression_summary}), with small and approximately symmetric exposure over- or under-predictions across income groups in most cities (e.g., Madrid, Barcelona and Bilbao; SI Fig.~\ref{fig:SI_E_income_per_city}). 

Age-related exposure differences are much smaller in magnitude (Fig.~\ref{fig:2}e). Thus, although regression slopes indicate that both models reproduce the sign and broad ordering of age-exposure differences (Fig.~\ref{fig:4}c and Fig.~\ref{fig:4}d), prediction errors become comparable to the magnitude of the empirical ones, limiting performance for both models (SI Table~\ref{tab:SI_gap_regression_summary}).

Together, these results suggest that group-agnostic mobility rules using population-based destination-choice mechanisms, combined with group-specific origin activity patterns, can reproduce an important component of aggregate heat exposure disparities across cities.

Beyond reproducing exposure differences, we further assessed whether the models capture structural properties of mobility networks across the urban heat field. Specifically, we examined heat assortativity and directionality of flows. Note that, even after removing intra-district trips, observed networks exhibit positive assortativity across both income and age groups, indicating a tendency for mobility flows to connect districts with similar temperature levels. In most of the cities, the gravity model systematically underestimates this assortative structure, while the radiation model tends to overestimate assortativity values (see SI Figs.~\ref{fig:SI_rho_income_per_city} and ~\ref{fig:SI_rho_age_per_city}). Related to directionality, both models generally reproduce the sign of heat-related directional components in mobility flows. However, in most cities neither model fully captures the magnitude of these directional effects, although the gravity model more closely matches their magnitude across groups than the radiation model (SI Fig.~\ref{fig:SI_R_income_per_city},~\ref{fig:SI_R_age_per_city}).

\section*{Discussion}

Our results reveal clear and systematic inequalities in heat exposure arising from daily mobility. Despite substantial heterogeneity in urban form and climatic conditions across the analyzed cities, the direction of inequality in mobility-based exposure is largely consistent: the low-income group shows higher mobility-based heat exposure than the high-income group ($\Delta E^{inc}>0$), and the younger group exhibits slightly higher exposure than the elderly group ($\Delta E^{age}>0$). Importantly, income emerges as the dominant axis of inequality in mobility-based heat exposure, while age-related differences are smaller and less consistent once stratified by income group. These disparities persist on a daily basis and are not driven by short-term fluctuations, indicating that they reflect structural differences in how demographic-group mobility flows are distributed across urban heat gradients. Trip-purpose analysis shows that commuting flows exhibit consistently higher exposure levels and stronger heat assortative mixing than non-routine trips, indicating that inequality is more pronounced for routine mobility. However, while non-routine mobility slightly attenuates exposure differences, it does not eliminate them, suggesting that inequality is structurally anchored in the broader spatial configuration of daily activity patterns rather than solely in work and education trips.

The comparison between home and non-home activities further clarifies the relationship between residential and mobility-based exposure estimates. Income- and age-related inequality magnitudes tend to be smaller for non-home destinations, although the sign is largely preserved, i.e. mobility does not change the residential ordering of group exposure. This pattern is consistent with the heat directionality results: mobility can shift groups toward hotter or cooler destinations relative to their origins, but these shifts tend to attenuate rather than reverse residence-based exposure differences. Thus, neglecting mobility-based exposure estimates may lead to incomplete estimates of heat exposure inequalities.

These findings extend residence-based evidence of heat inequality \cite[e.g.,][]{Hsu2021, Chakraborty2019} by showing that daily mobility patterns are associated with exposure disparities in a structurally persistent manner, consistent with prior work showing that mobility networks tend to connect socioeconomically similar areas \cite{Lenormand2023,Bokányi2021,Hilman2022}. Moreover, a key contribution of our framework is that it enables consistent cross-city comparison, allowing us to move beyond city-specific case studies to assess whether daily mobility maintains, attenuates, or amplifies residence-based exposure disparities across different urban contexts.

Our modeling results further support the structural interpretation of exposure inequality. Because model outflows are constrained to match the observed origin total trips for each group, exposure differences in the synthetic OD matrices reflect how common destination-choice mechanisms interact with group-specific origin outflow distributions and urban temperature gradients. Although the gravity model achieves higher aggregate OD overlap and lower mean absolute errors in group-specific exposure levels than the radiation model (SI Tables~\ref{tab:SI_cpc_slope_groups} and \ref{tab:SI_E_regression_summary}) -- consistent with previous comparative work \cite{Lenormand2016} -- it underestimates income- and age-related exposure differences. By contrast, the parameter-free radiation model captures most observed disparities, reflecting its ability to better preserve relative exposure contrasts between groups. This difference likely reflects how the models allocate destinations. The gravity model relies on a Newtonian attractiveness formulation, with flows determined by the pull between origin and destination populations modulated by a distance-deterrence function. By contrast, the radiation model allocates trips through intervening opportunities -- in our case, population -- so destinations are evaluated relative to the population located between origin and destination. These different destination-choice mechanisms produce distinct spatial flow structures (SI Figs.~\ref{fig:SI_flows_obs_modeled_income} and \ref{fig:SI_flows_obs_modeled_age}).

These results are consistent with the underlying spatial structure of cities. Across most cities, income and heat deciles are negatively associated ($\rho_{inc,heat}<0$; see SI Table~\ref{tab:structural_income_heat}), and both variables exhibit positive spatial autocorrelation (high Moran’s $I$; see SI Table~\ref{tab:structural_income_heat}), indicating clustered thermal and socioeconomic landscapes. In this context, groups originate from systematically different parts of the urban heat field, and mobility flows operate on top of this pre-existing spatial organization of daily activities. Consistently, the clearest reversals of income-related exposure inequality (i.e., higher exposure among high-income group $\Delta E^{inc}<0$) are observed in cities (e.g., Murcia, Zaragoza, and Oviedo) where the association between income and heat deciles is also reversed ($\rho_{inc,heat}>0$; see SI Table~\ref{tab:structural_income_heat}). Together, these results suggest that a substantial component of inequality in mobility-based exposure emerges from the interaction between residential segregation patterns, urban heat gradients, and group-agnostic mobility rules.

However, inequalities encoded by heat assortativity and directionality -- which depend not only on trip destinations but also on the interaction between origin and destination heat differences -- are not fully explained by generic trip allocation mechanisms common to the entire population. Although more complex models -- such as machine learning (ML) approaches \cite{Simini2021} or agent-based models \cite{lin2026modellinghumanactivitiescities} -- could improve predictive accuracy in city-specific applications, such extensions would come at the cost of generality and interpretability — key advantages of parsimonious mobility models \cite{Barthelemy2024}. These parsimonious formulations therefore provide useful benchmarks to identify cities with potentially high exposure inequalities warranting detailed intra-urban analysis. However, finer-scale planning applications will still require empirical mobility data or more detailed behavioral models, since our modeling does not identify the neighborhoods or routine destinations that contribute most to city-level inequalities, thereby leaving their identification as an important direction for follow-up work toward targeted heat-adaptation interventions.

With regard to methodology, we also note that our study has some limitations. First, although the main analysis focuses on Spanish cities, the external validation using US commuting data indicates that both the framework and the aggregated exposure-inequality patterns extend beyond the Spanish context. Nevertheless, broader validation across additional countries, mobility data sources, and climatic regions is still needed to assess how general these patterns of inequality in mobility-based heat exposure are. Second, mobile phone records identify activity labels and trips at the level of inferred visited locations, but do not provide the duration of each stay, the exact time spent outdoors, the transport mode, or the micro-environmental conditions experienced during the trip. Future work incorporating time-use information, transport mode, and indoor--outdoor conditions would help refine group-specific exposure estimates. Third, all analyses are conducted at the administrative district level. Because districts vary in size within and across cities, this aggregation may smooth local thermal heterogeneity and absorb short trips as intra-district movements, particularly in larger districts. However, since all sociodemographic groups within a city are evaluated under the same zoning system, this source of aggregation is less likely to mechanically generate the observed group differences and more likely to attenuate fine-scale exposure contrasts. Finally, the use of LST as a proxy for heat exposure requires careful consideration \cite{Zhao2025_LST_vs_airT}. LST may substantially differ from near-surface air temperature and thermal comfort indices, which are more robust indicators of human well-being and climate-related health risks \cite[e.g.,][]{Huang2023}. Yet, because our exposure metrics rely on relative rankings rather than absolute temperatures, our validation indicates that LST-based deciles preserve the spatial ordering of hotter and cooler areas also in terms of air temperature (see Methods). Future work incorporating high-resolution thermal comfort products \cite[e.g.,][]{roger2026hourly} would help refine exposure assessments and clarify the contexts in which LST is an adequate proxy. Another important direction for future research is to examine how extreme heat events reshape mobility patterns and associated mobility-based exposure inequalities. Our analysis focuses on structural mobility patterns and does not explicitly capture behavioral adaptations across sociodemographic groups during extreme heat events. Recent studies have shown that extreme heat reduces urban mobility, and that these reductions are heterogeneous across population groups \cite{Renninger2025, Stechemesser2023}. However, these studies do not examine how the spatial allocation of trips across hotter and cooler districts changes during extreme events, nor how such changes affect exposure inequalities. Understanding whether heatwaves mitigate or amplify existing mobility-based inequalities therefore remains an important direction for future research.

Overall, we provide a generalizable framework to characterize inequalities in mobility-based heat exposure. We show that these inequalities are structurally embedded features of cities, emerging from the interaction between heterogeneous urban heat fields and the socio-spatial organization of both residential locations and daily activity patterns. These results underscore the need for equity-aware urban planning that moves beyond residence-based assessments and considers how the spatial distribution of services, employment centers, and cooling infrastructure intersects with daily mobility patterns. As cities face intensifying heat extremes \cite{Vicedo-Cabrera2021}, understanding how daily mobility redistributes populations across urban temperature gradients is increasingly important for designing interventions that reduce climate-related inequalities.

\section*{Methods}

\subsection*{Data}

The main analysis focuses on 23 major Spanish cities (Fig.~\ref{fig:1}a), each with more than 200,000 inhabitants within the study boundaries. This threshold ensures sufficient spatial subdivision and mobility volume to robustly estimate intra-urban exposure inequalities. These cities encompass the largest urban areas in Spain while covering substantial heterogeneity in population size, climate, geography, transport infrastructure, and regional context. Municipalities adjoining a larger metropolitan core (e.g., Badalona in the Barcelona area) were excluded to avoid overlapping mobility basins.
Mobility and temperature datasets used in the main analysis correspond to July–August 2022 and July–August 2023. This period captures the peak of the warm season in Spain and provides sufficient temporal coverage to quantify mobility-based exposure patterns while minimizing missing data in both mobility and temperature fields.

Mobile phone mobility data were obtained from the Ministry of Transport, Mobility and Urban Agenda (MITMA), which provides daily origin–destination trip counts between Spanish administrative districts stratified by age-range ($0-25$, $25-45$, $45-65$, $65-100$ years), gender (female, male), and income class ($<10$, $10-15$, $>15$ thousands euros per year) \cite{mitma}. These data consist of aggregated counts of inferred trips rather than individual-level trajectories or persons. Trips are derived from sequences of inferred activity locations, and origin and destination are labeled according to the activity type (e.g., home, work/study, frequent, or infrequent activities). This allows us to distinguish trips originating or ending at home, which is central for distinguishing our mobility-based exposure estimates versus residential-based exposure results. For the analysis by trip purpose, we classify trips involving work or study activities as commuting trips, while trips involving infrequent activities are classified as non-routine trips, following the activity labels provided in the mobility dataset.

Temperature was represented using Land Surface Temperature (LST) products at 1 km resolution from NASA’s Moderate Resolution Imaging Spectroradiometer (MODIS) \cite{Wan2014}. In particular, we used the Terra LST 8-day composite product (MOD11A2) \cite{MOD11A2v61}, which provides temporally smoothed daytime LST fields by averaging all valid Terra observations within each 8-day window, reducing missing observations due to cloud contamination. For each district, temperature was computed as the average of all LST pixels within the district boundaries. Because some administrative districts are smaller than the MODIS pixel size, a small fraction of districts did not contain valid LST pixels. These districts were assigned temperature values through inverse-distance weighted interpolation based on neighboring districts.

To assess the robustness of LST-based exposure estimates, we further used the Aqua daily LST product (MYD11A1) \cite{MYD11A1v61}, which provides daytime LST, and 2m maximum air temperature from the UrbClim model \cite{temperature_urbclim} for July–August 2016 and 2017 in the subset of cities where these fields were available. Although UrbClim simulations correspond to a different temporal period than the mobility data, they are used solely to verify the spatial stability of intra-urban temperature rankings across datasets rather than to estimate exposure levels directly.

Demographic and socioeconomic attributes (median income, population) were obtained from the Spanish National Statistical Office (INE) \cite{ine}.

All data were aggregated to the administrative district level (Fig.~\ref{fig:1}b). City boundaries were defined using OpenStreetMap administrative polygons expanded by a uniform 1.2$\times$ buffer to approximate the functional urban area while preserving comparability across cities. Results are robust to alternative buffer sizes (see SI Fig.~\ref{fig:SI_boundary_test}). 

For external validation, we applied the same framework to 30 major US cities using tract-level home--work commuting flows from the Longitudinal Employer--Household Dynamics Origin--Destination Employment Statistics (LODES) database for 2022 \cite{LODES2024}. LODES provides commuting flows stratified by worker age and earnings categories, but not joint income--age information. Income groups were defined using the LODES earnings categories, comparing workers earning \$1,250 per month or less (SE01) with workers earning more than \$3,333 per month (SE03). Age groups were defined by comparing workers aged 29 years or younger (SA01) with workers aged 55 years or older (SA03).

Temperature was represented using daytime 2-m air-temperature estimates from the U-HAT dataset \cite{Milojevic2025UHAT}, which provides approximately 1-km resolution urban air-temperature estimates across major US metropolitan areas. We used the daytime prediction field, corresponding approximately to 1:30 PM local time, averaged over July--August 2022 and aggregated to Census tracts.

Because LODES captures home--work commuting flows rather than daily activity mobility, the US analysis should be interpreted as an external validation of the exposure framework using an independent commuting-based mobility dataset, rather than as a direct replication of the Spanish analysis.

\subsection*{Mobility matrices and exposure metrics}

\subsubsection*{Temperature deciles}

To compute mobility-based exposure metrics, continuous district-level temperature values were grouped into discrete classes. For each city, districts were ranked by their mean summer temperature during the study period (July-August 2022 and 2023), and partitioned into ten equal-frequency bins (deciles). This approach is widely used in socioeconomic segregation studies \cite{duransala2024,Bokányi2021,Hilman2022}, as it ensures that each bin contains a similar number of districts and allows comparisons across cities. By relying on relative thermal rankings rather than absolute temperature values, this classification captures the positioning of mobility flows within each city’s internal heat gradient, allowing consistent comparison of inequality patterns across cities with different climatic baselines. Moreover, using mean summer temperature to define deciles provides a stable heat field that avoids missing daily values and substantially reduces noise and computational burden. Using the median instead of the mean to define district temperatures yields nearly identical decile classifications and does not affect the resulting exposure inequalities. To evaluate the validity of this temporal aggregation, we performed two robustness checks (values are shown as mean~$\pm$~95\% CI):

(i) We compared the summer-averaged temperature aggregated deciles with daily temperature deciles computed independently for each city and day. Agreement is strong: across all city–day pairs, the quadratic-weighted $\kappa$ -- ranging from 0 to 1, with higher values indicating stronger agreement -- which quantifies agreement between two ordered classifications, assigning larger penalties to disagreements that span more deciles, was $0.80 \pm 0.01$. The mean absolute error (MAE) reflecting decile difference was $1.21 \pm 0.04$, with $28.6 \pm 1.1\%$ of districts shifting by more than one decile and $14.3 \pm 1.0\%$ shifting by more than two (see SI Fig.~\ref{fig:SI_monthly_vs_daily_deciles}). We further confirmed the agreement by recomputing the exposure inequalities using daily deciles for six randomly sampled days (see SI Fig.~\ref{fig:SI_random_days_test}), with results consistent with those obtained using aggregated deciles.

(ii) Because the main Spanish exposure estimates are computed from MODIS LST, we assessed whether LST-based deciles capture similar spatial patterns to near-surface air temperature deciles from UrbClim 2\,m maximum air temperature. In the subset of cities where both datasets were available (July–August 2016 and 2017), agreement between LST- and air-temperature-based deciles was substantial: the median quadratic-weighted kappa across cities was $0.73$. Thus, LST consistently reproduced the ranking of hotter and cooler districts across cities (see SI Fig.~\ref{fig:SI_LST_vs_air_deciles_Spain_US}). As an additional cross-dataset comparison, we also compared MODIS LST deciles with U-HAT 2\,m air-temperature deciles across the US cities. Agreement between LST- and air-temperature-based deciles was positive as well, with a median quadratic-weighted kappa of $0.51$ (see SI Fig.~\ref{fig:SI_LST_vs_air_deciles_Spain_US}).

\subsubsection*{Origin--destination (OD) matrices}

Let \(X_{ij}\) denote the number of trips originating in temperature decile \(i\) and ending in decile \(j\). To obtain the row-normalized conditional probability matrix \(C_{ij}\), we divided by the total trips originating from each decile \(i\):
\begin{equation}
C_{ij} = \frac{X_{ij}}{\sum_{j} X_{ij}}.
\end{equation}
Thus, each row of \(C\) sums to 1, and \(C_{ij}\) can be interpreted as a conditional probability of destination decile \(j\) given origin decile \(i\). Then, the marginal probability of originating in decile \(i\) is:
\begin{equation}
P(i) = \frac{\sum_{j} X_{ij}}{\sum_{i,j} X_{ij}},
\end{equation}
Using \(P(i)\) and \(C_{ij}\), we construct the fully normalized joint probability distribution \(\tilde{X}_{ij}\) as:
\begin{equation}
\tilde{X}_{ij} = P(i) C_{ij},
\end{equation}
where $\sum_{i,j} \tilde{X}_{ij} = 1$.

\subsubsection*{Mobility-based heat exposure}

Mobility-based heat exposure $E$ is defined as the trip-weighted mean temperature decile of visited destination districts:
\[
E = \frac{\sum_{i,j} j\,X_{ij}}{\sum_{i,j} X_{ij}}.
\]

This metric captures the average thermal level of the destinations reached by a given group through daily mobility. It is dimensionless and expressed in temperature-decile units. It is computed at the city level.

\subsubsection*{Heat exposure inequality}

To quantify mobility-based exposure inequalities, we defined the exposure difference for income groups ($\Delta E^{inc}$) and age groups ($\Delta E^{age}$) as:

\begin{equation}
\Delta E^{\text{inc}} = E^{<10} - E^{>15}, 
\qquad
\Delta E^{\text{age}} = E^{0\text{–}25} - E^{65\text{–}100}.
\end{equation}

where positive values of $\Delta E$ indicate higher exposure for low-income (or younger) group relative to high-income (or elderly) group within the same city.

\subsubsection*{Heat assortativity}

Heat assortativity quantifies whether trips connect districts of similar temperature rank. Following \cite{Bokányi2021}, the Pearson assortativity is defined as

\begin{equation}
\rho = \frac{\sum_{i,j}ij \tilde{X}_{ij} - \sum_{i,j}i\tilde{X}_{ij}\sum_{i,j} j \tilde{X}_{ij}}
{
\sqrt{\sum_{i,j} i^2 \tilde{X}_{ij} - (\sum_{i,j} i \tilde{X}_{ij})^2}
\sqrt{\sum_{i, j} j^2\tilde{X}_{ij} - (\sum_{i,j} j \tilde{X}_{ij})^2}},
\quad -1 \le \rho \le 1.
\end{equation}
Positive values of \(\rho\) indicate assortative mobility, meaning that trips occur between districts with similar temperature deciles, while negative values indicate disassortative mobility, with trips occurring between dissimilar districts. In this context, assortativity captures the extent to which mobility reinforces or mitigates existing residential thermal contrasts within cities.

\subsubsection*{Heat directionality}

Heat directionality measures whether trips tend to move toward hotter or cooler areas. Let
\[
U = \sum_{i<j} X_{ij} \quad\text{and}\quad
L = \sum_{i>j} X_{ij},
\]
denote the number of trips to higher- and lower-temperature deciles, respectively. The directionality index is defined as
\[
R = \frac{U - L}{U + L},
\]
with \(R=1\) meaning all trips move toward hotter deciles, \(R=-1\) toward cooler ones, and \(R=0\) indicating balance. This metric captures the net thermal bias of mobility flows across the urban heat field, quantifying whether daily movement patterns systematically shift populations toward hotter or cooler environments relative to their origins.

\subsubsection*{Population-weighted average metric}

To summarize metric patterns at the national scale, we computed population-weighted averages of the different metrics as:

\[
\bar{M}= 
\frac{\sum_{c} P_c \, M_{c}}
{\sum_{c} P_c}.
\]

where $M_{c}$ denotes the heat mobility-based metric -- i.e., exposure, assortativity, or directionality -- in city $c$, and $P_c$ denotes the total population of city $c$ within the defined urban boundary. This weighting ensures that each city contributes proportionally to its demographic relevance, so that $\bar{M}$ represents the value of the metric experienced by the average urban resident in Spain.

\subsection*{Population-based mobility models}

To assess whether standard mobility models can reproduce the spatial patterns of heat-exposure inequalities observed in empirical flows, we implemented two population-based models: the gravity and radiation models. Our objective is not to reproduce the absolute number of trips generated by each district, but rather to assess whether these population-based formulations can recover the spatial allocation of flows across districts and, consequently, the relative exposure differences between social groups. This motivated two implementation choices: (i) an origin-constrained form, such that the total outflow from each district \(i\) matches the observed outflow $\sum_j T_{ij}^{\mathrm{model}} = O_i^{\mathrm{obs}}$, and (ii) removal of intra-district flows. All comparisons therefore concern how models redistribute flows between districts, conditional on the observed outflows. For comparisons between empirical and modeled metrics, empirical OD matrices were recomputed after excluding intra-district flows, to match the model specification.

\subsubsection*{Gravity model}
We implemented a classical Poisson gravity model \cite{Zipf1946} in which expected flows between districts \(i\) and \(j\) scale proportionally with origin and destination populations and decay with distance as
\[
T_{ij}^{\mathrm{grav}} \propto d_{ij}^{\beta_1} P_i^{\beta_2} P_j^{\beta_3},
\]
where \(P_i\) and \(P_j\) denote the residential populations of the origin and destination districts, and \(d_{ij}\) is the centroid-to-centroid distance. Parameters \((\beta_0, \beta_1,\beta_2,\beta_3)\) are estimated via Poisson regression using log-distance and log-populations as predictors:
\[
\log \mathbb{E}[T_{ij}^{\mathrm{grav}}] = \beta_0 + \beta_1 \log d_{ij} + \beta_2 \log P_i + \beta_3 \log P_j.
\]
To keep the gravity model parsimonious and group-agnostic, we estimated a single parameter vector \((\beta_0,\beta_1,\beta_2,\beta_3)\) using total inter-district flows pooled across cities and sociodemographic groups. The same fitted parameters were then applied to every city and group. For each city, group, and origin district, predicted flows were rescaled to match the observed origin outflow $\sum_{j\neq i} T_{ij}^{grav} = O_i^{\mathrm{obs}}$. Intra-district flows are excluded by setting \(T_{ii}=0\).

\subsubsection*{Radiation model}

The radiation model \citep{Simini2012} is a parameter-free model, based on intervening opportunities. For each pair of districts \(i\) and \(j\), the expected number of trips is

\[
T_{ij}^{\mathrm{rad}} = O_i^{\mathrm{obs}} \,
\frac{P_i P_j}{(P_i + s_{ij})(P_i + P_j + s_{ij})}.
\]

where \(P_i\) and \(P_j\) denote the residential populations of the origin and destination districts, and \(s_{ij}\) is the total residential population within a circle centered at \(i\) with radius \(d_{ij}\), excluding \(P_i\) and \(P_j\). This formulation is origin-constrained by construction. For each city we pre-computed the full radiation kernel using district centroids and residential populations, and generated synthetic OD matrices for each income and age group by applying the corresponding group-specific origin outflows and removing all diagonal entries (\(T_{ii}=0\)) to isolate inter-district mobility. Note that the radiation kernel is identical for all income and age groups: differences in predicted flows arise solely from group-specific origin total trips, not from any group-specific mobility parameters.

\subsubsection*{Evaluation of model performance for exposure inequalities}

For each city $c$, we computed the observed and model-predicted exposure differences using the radiation and gravity models. We then estimated an ordinary least squares regression of predicted differences against observed ones
\[
\Delta E_{\mathrm{pred},c} = a + b \, \Delta E_{\mathrm{obs},c}
\]
where the slope $b$ quantifies the proportional agreement between modeled and observed inequalities, while the intercept $a$ captures displacement when $\Delta E^{\mathrm{obs}} = 0$. We also report the coefficient of determination ($R^2$), the bias defined as the average predicted minus observed exposure difference across cities, and the mean absolute error (MAE) defined as the average absolute difference between predicted and observed exposure differences across cities (SI Table~\ref{tab:SI_gap_regression_summary}).

We further evaluated how well the models reproduce absolute exposure levels $E$ for each group by regressing predicted exposure values against the corresponding observed values across cities (SI Table~\ref{tab:SI_E_regression_summary}).

\section*{Figures and Tables}

\section*{Acknowledgements}
G.M. acknowledges support from the SNSF Weave/Lead Agency funding scheme (grant number 213995).

\section*{Author Contributions}
G.M. and M.D.S. conceived the study. M.D.S. performed the analysis and drafted the manuscript. All authors interpreted the results, provided feedback that helped shape the analysis, and contributed to writing the manuscript.

\section*{Competing Interests}
The authors declare no competing interests.

\section*{Data and Code Availability}

The Jupyter notebooks and associated scripts necessary to reproduce the results of this paper are accessible on GitHub at \url{https://github.com/urbes-team/mobility_based_heat_exposure_inequality}.

\bibliographystyle{unsrtnat}  
\bibliography{bib}            

\newpage
\section*{Supplementary Information to: \\ ''Mobility shapes heat exposure inequalities in cities''}

{\large Marc Duran-Sala$^{1,*}$, Mattia Mazzoli$^2$, Martin Hendrick$^{1}$, Gabriele Manoli$^{1,*}$} \\
$^{1}$Laboratory of Urban and Environmental Systems, \'{E}cole Polytechnique F\'{e}d\'{e}rale de Lausanne, Lausanne, Switzerland \\
$^{2}$ ISI Foundation, Turin, Italy\\
$^{*}$Corresponding author(s): marc.duransala@epfl.ch, gabriele.manoli@epfl.ch






\newcommand{\beginsupplement}{%
        \setcounter{table}{0}
        \renewcommand{\thetable}{S.\arabic{table}}%
        \setcounter{figure}{0}
        \renewcommand{\thefigure}{S.\arabic{figure}}%
        \setcounter{equation}{0}
        \renewcommand{\theequation}{S.\arabic{equation}}%
        \setcounter{section}{0}
        \renewcommand{\thesection}{S.\arabic{section}}%
     }

 \beginsupplement

\section*{Supplementary Figures and Tables}

\begin{table}[H]
\centering
\caption{Temperature (LST) summary for Spanish cities}
\vspace{1em}

\label{tab:temp_cities}
\begin{tabular}{lrrr}
city & Mean (°C) & Min (°C) & Max (°C) \\
\midrule
A\_Coruna & 29.76 & 22.83 & 33.53 \\
Alicante & 40.01 & 36.30 & 42.15 \\
Barcelona & 37.92 & 33.16 & 41.80 \\
Bilbao & 32.93 & 29.06 & 37.50 \\
Cartagena & 40.86 & 37.75 & 44.18 \\
Cordoba & 47.76 & 43.29 & 51.49 \\
Elche & 40.82 & 35.71 & 42.74 \\
Gijon & 31.05 & 27.25 & 34.02 \\
Granada & 45.83 & 38.80 & 48.71 \\
Jerez\_de\_la\_Frontera & 41.36 & 30.62 & 50.45 \\
Las\_Palmas & 34.40 & 31.09 & 39.65 \\
Madrid & 45.80 & 31.70 & 50.00 \\
Malaga & 40.75 & 37.51 & 45.55 \\
Murcia & 42.38 & 38.89 & 44.43 \\
Oviedo & 31.13 & 27.21 & 36.01 \\
Palma\_de\_Mallorca & 39.30 & 31.90 & 43.46 \\
Santa\_Cruz\_de\_Tenerife & 37.06 & 29.37 & 40.83 \\
Sevilla & 47.86 & 44.57 & 51.17 \\
Valencia & 39.68 & 29.91 & 42.06 \\
Valladolid & 43.58 & 41.41 & 45.53 \\
Vigo & 29.83 & 26.18 & 33.19 \\
Vitoria\_Gasteiz & 37.81 & 28.92 & 39.99 \\
Zaragoza & 42.48 & 38.54 & 45.65 \\
\end{tabular}
\label{tab:SI_cities_summary}
\end{table}

\begin{table}[H]
\centering
\caption{
Air temperature summary for US cities.}
\vspace{1em}
\begin{tabular}{lrrr}
city & Mean (°C) & Min (°C) & Max (°C) \\
\midrule
Chicago & 27.25 & 22.63 & 28.69 \\
Boston & 29.34 & 25.82 & 30.63 \\
Los\_Angeles & 30.00 & 25.00 & 33.02 \\
San\_Jose & 26.83 & 23.89 & 29.00 \\
Oakland & 25.16 & 23.32 & 26.99 \\
Minneapolis & 27.17 & 23.68 & 28.28 \\
Philadelphia & 31.14 & 28.70 & 33.08 \\
Seattle & 25.59 & 20.22 & 26.91 \\
Baltimore & 29.05 & 25.53 & 30.25 \\
Portland & 28.76 & 26.13 & 29.87 \\
San\_Diego & 27.35 & 23.26 & 30.47 \\
Jacksonville & 31.68 & 30.28 & 32.72 \\
Milwaukee & 25.98 & 24.33 & 27.17 \\
Miami & 31.44 & 30.07 & 32.28 \\
NYC & 29.81 & 25.58 & 31.31 \\
Houston & 35.20 & 32.75 & 37.14 \\
Atlanta & 30.56 & 28.59 & 31.27 \\
Sacramento & 33.73 & 32.61 & 34.67 \\
Nashville & 31.01 & 30.04 & 32.94 \\
Charlotte & 31.08 & 30.11 & 32.69 \\
Denver & 34.24 & 31.04 & 35.70 \\
Austin & 35.92 & 34.78 & 36.93 \\
Washington\_DC & 29.45 & 27.32 & 31.50 \\
San\_Antonio & 36.12 & 34.69 & 37.22 \\
Memphis & 33.34 & 31.27 & 34.47 \\
San\_Francisco & 23.09 & 19.56 & 26.02 \\
Las\_Vegas & 37.48 & 35.95 & 39.21 \\
Columbus & 28.38 & 27.32 & 29.37 \\
Phoenix & 38.41 & 36.12 & 39.76 \\
Dallas & 36.44 & 34.84 & 37.83 \\
\end{tabular}
\label{tab:SI_US_cities_summary}
\end{table}

\begin{table}[H]
\centering
\caption{
Paired city-level statistical tests for heat exposure inequalities by joint income--age stratifications across Spanish cities. We report mean exposure differences across cities, the number of cities with positive and negative differences, and two-sided Wilcoxon signed-rank and sign test p-values. For income, $\Delta E^{inc}=E_{<10}-E_{>15}$, and for age, $\Delta E^{age}=E_{young}-E_{older}$. Here, $\langle \cdot \rangle$ denotes the average across cities.
}
\vspace{1em}

\textbf{(A) Low- and high-income exposure differences} \\
\vspace{0.35em}
\begin{tabular}{lcccc}
Age group & $\langle \Delta E^{inc} \rangle$ & $n_+/n_-$ & Wilcoxon $p$ & Sign $p$ \\
\midrule
All (marginal) & 0.58 & 17/6 & $1.23\times10^{-2}$ & $3.47\times10^{-2}$ \\
0--25 & 0.57 & 17/6 & $1.32\times10^{-2}$ & $3.47\times10^{-2}$ \\
25--45 & 0.58 & 17/6 & $1.12\times10^{-2}$ & $3.47\times10^{-2}$ \\
45--65 & 0.55 & 17/6 & $1.63\times10^{-2}$ & $3.47\times10^{-2}$ \\
65--100 & 0.63 & 19/4 & $4.85\times10^{-3}$ & $2.60\times10^{-3}$ \\
\end{tabular}

\vspace{1.2em}

\textbf{(B) Young and elderly exposure differences} \\
\vspace{0.35em}
\begin{tabular}{lcccc}
Income group & $\langle \Delta E^{age} \rangle$ & $n_+/n_-$ & Wilcoxon $p$ & Sign $p$ \\
\midrule
All (marginal) & 0.14 & 19/4 & $8.85\times10^{-5}$ & $2.60\times10^{-3}$ \\
$<10$ & 0.05 & 13/10 & $6.05\times10^{-1}$ & $6.78\times10^{-1}$ \\
10--15 & 0.08 & 16/7 & $3.09\times10^{-2}$ & $9.31\times10^{-2}$ \\
$>15$ & 0.11 & 19/4 & $3.45\times10^{-3}$ & $2.60\times10^{-3}$ \\
\end{tabular}
\label{tab:SI_paired_joint_combined}
\end{table}

\begin{table}[H]
\centering
\caption{
Paired city-level statistical tests for heat exposure inequalities across US cities. We report mean exposure differences across cities, the number of cities with positive and negative differences, and two-sided Wilcoxon signed-rank and sign test p-values. For income, $\Delta E^{inc}=E_{low}-E_{high}$, and for age, $\Delta E^{age}=E_{young}-E_{older}$. Here, $\langle \cdot \rangle$ denotes the average across cities.
}
\vspace{1em}

\textbf{(A) Low- and high-income exposure differences} \\
\vspace{0.35em}
\begin{tabular}{lcccc}
Group & $\langle \Delta E^{inc} \rangle$ & $n_+/n_-$ & Wilcoxon $p$ & Sign $p$ \\
\midrule
All (marginal) & 0.28 & 18/12 & $1.13\times10^{-2}$ & $0.36$ \\
\end{tabular}

\vspace{1.2em}

\textbf{(B) Young and older exposure differences} \\
\vspace{0.35em}
\begin{tabular}{lcccc}
Group & $\langle \Delta E^{age} \rangle$ & $n_+/n_-$ & Wilcoxon $p$ & Sign $p$ \\
\midrule
All (marginal) & -0.06 & 10/20 & $6.12\times10^{-2}$ & $0.10$ \\
\end{tabular}

\label{tab:SI_US_paired_tests}
\end{table}

\begin{table}[H]
\centering
\renewcommand{\arraystretch}{0.9}
\caption{
City-level mobility heat metrics computed from mobility flows aggregated over the July--August 2022 and 2023 study period. For each city, we show the heat exposure ($E$), heat assortativity ($\rho$), and heat directionality ($R$) for income groups (low-income $<10$ vs high-income $>15$) and age groups (young 0--25 vs elderly 65--100).
}
\vspace{0.7em}

\textbf{Income groups}\\
\vspace{0.35em}
\begin{tabular}{lcccccc}
City &
\multicolumn{3}{c}{Low ($<10$)} &
\multicolumn{3}{c}{High ($>15$)} \\
& $E$ & $\rho$ & $R$ & $E$ & $\rho$ & $R$ \\
\midrule
Madrid & 7.32 & 0.54 & -0.16 & 3.92 & 0.58 & 0.02 \\
Barcelona & 5.42 & 0.29 & 0.15 & 4.70 & 0.27 & 0.11 \\
Valencia & 5.94 & 0.41 & 0.36 & 6.14 & 0.42 & -0.02 \\
Zaragoza & 5.31 & 0.29 & -0.03 & 5.79 & 0.22 & -0.17 \\
Sevilla & 5.48 & 0.39 & -0.06 & 3.73 & 0.40 & 0.33 \\
Malaga & 3.99 & 0.54 & -0.06 & 3.84 & 0.31 & 0.09 \\
Murcia & 5.59 & 0.58 & -0.06 & 6.90 & 0.63 & -0.08 \\
Palma de Mallorca & 4.49 & 0.35 & 0.03 & 4.26 & 0.32 & 0.28 \\
Las Palmas & 5.62 & 0.30 & -0.39 & 4.83 & 0.39 & -0.40 \\
Alicante & 5.67 & 0.29 & 0.07 & 4.68 & 0.21 & 0.36 \\
Bilbao & 5.78 & 0.05 & -0.00 & 5.78 & 0.17 & -0.04 \\
Valladolid & 5.63 & 0.18 & 0.01 & 5.58 & 0.06 & 0.15 \\
Cordoba & 5.71 & 0.36 & -0.22 & 4.66 & 0.43 & -0.00 \\
Vigo & 6.26 & 0.27 & 0.25 & 5.88 & 0.43 & 0.04 \\
Gijon & 5.28 & 0.26 & 0.02 & 5.59 & 0.06 & 0.20 \\
Vitoria-Gasteiz & 8.81 & 0.58 & -0.05 & 7.97 & 0.57 & -0.22 \\
A Coruna & 4.60 & 0.26 & 0.03 & 4.86 & 0.22 & 0.26 \\
Granada & 5.97 & 0.17 & -0.03 & 5.04 & 0.13 & 0.22 \\
Elche & 4.88 & 0.57 & -0.28 & 2.81 & 0.63 & 0.02 \\
Cartagena & 4.89 & 0.58 & -0.02 & 4.25 & 0.41 & 0.11 \\
Oviedo & 6.14 & 0.49 & -0.01 & 6.50 & 0.26 & 0.13 \\
Jerez de la Frontera & 5.47 & 0.59 & -0.10 & 4.20 & 0.45 & 0.16 \\
Santa Cruz de Tenerife & 7.60 & 0.35 & -0.19 & 6.57 & 0.16 & 0.09 \\
\end{tabular}

\vspace{1.2em}

\textbf{Age groups}\\
\vspace{0.35em}
\begin{tabular}{lcccccc}
City &
\multicolumn{3}{c}{Young (0--25)} &
\multicolumn{3}{c}{Elderly (65--100)} \\
& $E$ & $\rho$ & $R$ & $E$ & $\rho$ & $R$ \\
\midrule
Madrid & 4.99 & 0.58 & -0.05 & 4.66 & 0.61 & 0.01 \\
Barcelona & 5.39 & 0.34 & 0.01 & 5.37 & 0.39 & 0.01 \\
Valencia & 5.97 & 0.49 & -0.04 & 5.95 & 0.56 & -0.04 \\
Zaragoza & 5.54 & 0.25 & -0.00 & 5.46 & 0.26 & 0.05 \\
Sevilla & 5.22 & 0.49 & -0.08 & 4.95 & 0.54 & -0.03 \\
Malaga & 4.24 & 0.50 & -0.05 & 3.97 & 0.50 & 0.02 \\
Murcia & 6.45 & 0.49 & 0.02 & 6.16 & 0.58 & 0.01 \\
Palma de Mallorca & 4.77 & 0.40 & -0.03 & 4.76 & 0.39 & -0.01 \\
Las Palmas & 4.28 & 0.37 & -0.13 & 4.02 & 0.42 & -0.08 \\
Alicante & 5.89 & 0.42 & -0.08 & 5.78 & 0.45 & -0.02 \\
Bilbao & 5.73 & 0.09 & 0.07 & 5.79 & 0.10 & -0.00 \\
Valladolid & 5.68 & 0.18 & -0.03 & 5.70 & 0.22 & 0.01 \\
Cordoba & 4.97 & 0.36 & 0.05 & 4.81 & 0.35 & 0.03 \\
Vigo & 6.45 & 0.44 & -0.03 & 6.43 & 0.47 & -0.03 \\
Gijon & 5.49 & 0.23 & -0.18 & 5.50 & 0.26 & -0.17 \\
Vitoria-Gasteiz & 8.13 & 0.54 & -0.22 & 8.17 & 0.59 & -0.12 \\
A Coruna & 5.17 & 0.33 & 0.02 & 4.94 & 0.35 & 0.03 \\
Granada & 5.60 & 0.20 & -0.03 & 5.44 & 0.20 & -0.02 \\
Elche & 4.89 & 0.54 & -0.17 & 4.48 & 0.63 & -0.10 \\
Cartagena & 4.88 & 0.51 & -0.09 & 4.55 & 0.52 & -0.07 \\
Oviedo & 6.14 & 0.56 & 0.06 & 5.92 & 0.62 & 0.08 \\
Jerez de la Frontera & 4.86 & 0.64 & -0.01 & 4.60 & 0.70 & 0.03 \\
Santa Cruz de Tenerife & 6.94 & 0.37 & -0.04 & 6.84 & 0.37 & -0.01 \\
\end{tabular}
\label{tab:SI_city_metrics_income_age}
\end{table}

\begin{table}[H]
\centering
\setlength{\tabcolsep}{10pt}
\caption{
Bootstrap-based robustness analysis of heat exposure differences. For each city, we report the bootstrap mean and 95\% confidence interval of the exposure difference between: (A) low- and high-income groups ($\Delta E^{inc}$) and (B) young and elderly population groups ($\Delta E^{age}$).
Bootstrap resampling was performed using a multinomial procedure with $n_{\mathrm{boot}} = 1000$ replicates. For each bootstrap replicate, we redistributed the observed aggregated trip counts across OD pairs according to their empirical probabilities while keeping the total number of trips per city and group fixed.
All reported estimates are statistically significant based on a two-sided bootstrap test ($p < 0.001$), denoted by $^{***}$.
}
\vspace{1.2em}
\begin{tabular}{l@{\hspace{20pt}}l@{\hspace{20pt}}l}
City &
\multicolumn{1}{c}{$\Delta E^{inc}$} &
\multicolumn{1}{c}{$\Delta E^{age}$} \\
\midrule
Madrid & 3.41$^{***}$ [3.41, 3.41] & 0.33$^{***}$ [0.33, 0.33] \\
Barcelona & 0.71$^{***}$ [0.71, 0.71] & 0.03$^{***}$ [0.03, 0.03] \\
Valencia & -0.20$^{***}$ [-0.20, -0.20] & 0.03$^{***}$ [0.02, 0.03] \\
Zaragoza & -0.48$^{***}$ [-0.49, -0.48] & 0.08$^{***}$ [0.08, 0.08] \\
Sevilla & 1.76$^{***}$ [1.76, 1.76] & 0.27$^{***}$ [0.27, 0.27] \\
Malaga & 0.16$^{***}$ [0.15, 0.16] & 0.27$^{***}$ [0.26, 0.27] \\
Murcia & -1.31$^{***}$ [-1.32, -1.31] & 0.29$^{***}$ [0.29, 0.29] \\
Palma de Mallorca & 0.23$^{***}$ [0.22, 0.23] & 0.01$^{***}$ [0.01, 0.01] \\
Las Palmas & 0.79$^{***}$ [0.79, 0.79] & 0.26$^{***}$ [0.26, 0.26] \\
Alicante & 0.98$^{***}$ [0.98, 0.98] & 0.11$^{***}$ [0.11, 0.11] \\
Bilbao & 0.00$^{***}$ [0.00, 0.01] & -0.06$^{***}$ [-0.06, -0.06] \\
Valladolid & 0.05$^{***}$ [0.05, 0.05] & -0.02$^{***}$ [-0.02, -0.02] \\
Cordoba & 1.05$^{***}$ [1.05, 1.06] & 0.17$^{***}$ [0.16, 0.17] \\
Vigo & 0.38$^{***}$ [0.38, 0.38] & 0.02$^{***}$ [0.02, 0.02] \\
Gijon & -0.31$^{***}$ [-0.31, -0.30] & -0.01$^{***}$ [-0.01, -0.01] \\
Vitoria-Gasteiz & 0.84$^{***}$ [0.84, 0.84] & -0.04$^{***}$ [-0.05, -0.04] \\
A Coruna & -0.26$^{***}$ [-0.27, -0.25] & 0.23$^{***}$ [0.23, 0.23] \\
Granada & 0.93$^{***}$ [0.93, 0.93] & 0.16$^{***}$ [0.16, 0.16] \\
Elche & 2.07$^{***}$ [2.06, 2.07] & 0.41$^{***}$ [0.41, 0.41] \\
Cartagena & 0.64$^{***}$ [0.64, 0.64] & 0.33$^{***}$ [0.33, 0.33] \\
Oviedo & -0.36$^{***}$ [-0.37, -0.35] & 0.23$^{***}$ [0.22, 0.23] \\
Jerez de la Frontera & 1.27$^{***}$ [1.27, 1.27] & 0.26$^{***}$ [0.26, 0.26] \\
Santa Cruz de Tenerife & 1.03$^{***}$ [1.02, 1.03] & 0.10$^{***}$ [0.10, 0.10] \\
\end{tabular}
\label{tab:SI_bootstrap}
\end{table}


\begin{table}[H]
\centering
\caption{District level structural association between income and heat across cities. 
Spearman $\rho$ measures the rank correlation between income and heat deciles. Spearman’s $\rho$ ranges from $-1$ (perfect negative rank association) to $1$ (perfect positive rank association), with $0$ indicating no rank relationship. Moran's $I$ is computed on deciles using KNN weights.}
\vspace{1em}

\begin{tabular}{lrrr}
City & $\rho_{\text{inc,heat}}$ & $I_{\text{heat}}$ & $I_{\text{income}}$ \\
\midrule
A\_Coruna & -0.38 & 0.59 & 0.25 \\
Alicante & -0.06 & 0.58 & 0.08 \\
Barcelona & -0.28 & 0.53 & 0.37 \\
Bilbao & -0.30 & 0.23 & 0.29 \\
Cartagena & -0.10 & 0.45 & 0.10 \\
Cordoba & 0.03 & 0.66 & 0.48 \\
Elche & -0.04 & 0.70 & 0.23 \\
Gijon & -0.31 & 0.34 & -0.09 \\
Granada & -0.28 & 0.45 & 0.30 \\
Jerez\_de\_la\_Frontera & -0.49 & 0.75 & 0.35 \\
Las\_Palmas & -0.44 & 0.52 & -0.12 \\
Madrid & -0.32 & 0.69 & 0.45 \\
Malaga & -0.43 & 0.77 & 0.50 \\
Murcia & 0.50 & 0.55 & 0.39 \\
Oviedo & 0.21 & 0.73 & 0.58 \\
Palma\_de\_Mallorca & -0.46 & 0.65 & 0.54 \\
Santa\_Cruz\_de\_Tenerife & -0.18 & 0.56 & 0.22 \\
Sevilla & -0.08 & 0.37 & 0.15 \\
Valencia & -0.09 & 0.74 & 0.49 \\
Valladolid & -0.11 & 0.45 & 0.48 \\
Vigo & -0.04 & 0.69 & 0.44 \\
Vitoria\_Gasteiz & -0.62 & 0.65 & 0.43 \\
Zaragoza & 0.17 & 0.56 & 0.40 \\
\end{tabular}

\label{tab:structural_income_heat}
\end{table}

\begin{table}[H]
\centering
\caption{Regression-based comparison between observed and predicted exposure differences across cities. For each model, we regress the predicted exposure difference ($\Delta E^{age}$ or $\Delta E^{inc}$) against the corresponding observed difference (one point per city; $n=23$). We report the slope ($b$), intercept ($a$), coefficient of determination ($R^2$), the bias defined as average predicted minus observed exposure difference across cities, and the mean absolute error (MAE), defined as the average absolute difference between predicted and observed exposure differences across cities.}
\vspace{0.8em}

\textbf{(A) Income exposure difference}\\
\begin{tabular}{lccccc}
Model & Slope, $b$ & Intercept, $a$ & Bias & MAE & $R^2$ \\
\midrule
Radiation & 0.87 & -0.04 & -0.09 & 0.29 & 0.78 \\
Gravity   & 0.45 & -0.04 & -0.28 & 0.41 & 0.75 \\
\end{tabular}

\vspace{1.1em}

\textbf{(B) Age exposure difference}\\
\begin{tabular}{lccccc}
Model & Slope, $b$ & Intercept, $a$ & Bias & MAE & $R^2$ \\
\midrule
Radiation & 1.08 & 0.02  & 0.03  & 0.08 & 0.56 \\
Gravity   & 0.57 & -0.01 & -0.05 & 0.07 & 0.64 \\
\end{tabular}

\label{tab:SI_gap_regression_summary}
\end{table}

\begin{table}[H]
\centering
\caption{Regression-based comparison between observed and predicted group-specific heat exposure levels across cities. For each model, we regress the predicted exposure $E$ against the corresponding observed exposure for each sociodemographic group (one point per city; $n=23$). We report the slope ($b$), intercept ($a$), coefficient of determination ($R^2$), the bias defined as average predicted minus observed exposure across cities, and the mean absolute error (MAE), defined as the average absolute difference between predicted and observed exposure values across cities.}
\vspace{0.8em}

\textbf{(A) Income}\\
\begin{tabular}{llccccc}
Group & Model & Slope, $b$ & Intercept, $a$ & Bias & MAE & $R^2$ \\
\midrule
$<10$   & Gravity   & 0.82 & 1.18  & 0.14 & 0.40 & 0.67 \\
$<10$   & Radiation & 0.94 & 0.60  & 0.26 & 0.45 & 0.71 \\
10--15  & Gravity   & 0.94 & 0.50  & 0.17 & 0.37 & 0.76 \\
10--15  & Radiation & 1.08 & -0.15 & 0.28 & 0.55 & 0.72 \\
$>15$   & Gravity   & 0.83 & 1.30  & 0.42 & 0.53 & 0.76 \\
$>15$   & Radiation & 1.01 & 0.31  & 0.36 & 0.54 & 0.74 \\
\end{tabular}

\vspace{1.1em}

\textbf{(B) Age}\\
\begin{tabular}{llccccc}
Group & Model & Slope, $b$ & Intercept, $a$ & Bias & MAE & $R^2$ \\
\midrule
0--25    & Gravity   & 0.99 & 0.24  & 0.18 & 0.33 & 0.77 \\
0--25    & Radiation & 1.08 & -0.17 & 0.30 & 0.51 & 0.73 \\
25--45   & Gravity   & 0.98 & 0.32  & 0.18 & 0.33 & 0.78 \\
25--45   & Radiation & 1.08 & -0.17 & 0.28 & 0.50 & 0.73 \\
45--65   & Gravity   & 0.97 & 0.38  & 0.20 & 0.34 & 0.78 \\
45--65   & Radiation & 1.07 & -0.13 & 0.29 & 0.50 & 0.74 \\
65--100  & Gravity   & 0.96 & 0.45  & 0.24 & 0.36 & 0.79 \\
65--100  & Radiation & 1.09 & -0.21 & 0.27 & 0.50 & 0.75 \\
\end{tabular}

\label{tab:SI_E_regression_summary}
\end{table}

\begin{table}[H]
\centering
\caption{Summary for (A) income groups and (B) age groups, using gravity and radiation models of (i) the Common Part of Commuters (CPC), a Sørensen–Dice similarity index widely used in mobility research that quantifies the fraction of OD flows correctly reproduced by the model (CPC = 1 indicates perfect overlap), and (ii) the slope obtained from regressing modeled travel probabilities against observed ones. Values are shown as mean~$\pm$~95\% confidence intervals across cities.}
\vspace{1em}

\textbf{(A) Income groups}\\
\begin{tabular}{llcc}
Group & Model & CPC & Slope \\
\midrule
$<10$  & Radiation & 0.32 $\pm$ 0.01 & 0.99 $\pm$ 0.02 \\
$<10$  & Gravity   & 0.39 $\pm$ 0.01 & 0.63 $\pm$ 0.03 \\
10--15 & Radiation & 0.34 $\pm$ 0.01 & 1.02 $\pm$ 0.01 \\
10--15 & Gravity   & 0.39 $\pm$ 0.00 & 0.65 $\pm$ 0.03 \\
$>15$  & Radiation & 0.33 $\pm$ 0.01 & 1.02 $\pm$ 0.01 \\
$>15$  & Gravity   & 0.38 $\pm$ 0.01 & 0.69 $\pm$ 0.03 \\
\end{tabular}

\vspace{1.2em}

\textbf{(B) Age groups}\\
\begin{tabular}{llcc}
Group & Model & CPC & Slope \\
\midrule
0--25   & Radiation & 0.33 $\pm$ 0.01 & 0.90 $\pm$ 0.03 \\
0--25   & Gravity   & 0.45 $\pm$ 0.02 & 0.38 $\pm$ 0.04 \\
25--45  & Radiation & 0.33 $\pm$ 0.01 & 0.89 $\pm$ 0.04 \\
25--45  & Gravity   & 0.45 $\pm$ 0.02 & 0.40 $\pm$ 0.04 \\
45--65  & Radiation & 0.33 $\pm$ 0.01 & 0.89 $\pm$ 0.04 \\
45--65  & Gravity   & 0.45 $\pm$ 0.02 & 0.38 $\pm$ 0.04 \\
65--100 & Radiation & 0.34 $\pm$ 0.01 & 0.91 $\pm$ 0.04 \\
65--100 & Gravity   & 0.43 $\pm$ 0.02 & 0.42 $\pm$ 0.05 \\
\end{tabular}

\label{tab:SI_cpc_slope_groups}
\end{table}



\begin{figure}[H]   
    \centering
    \includegraphics[width=\linewidth]{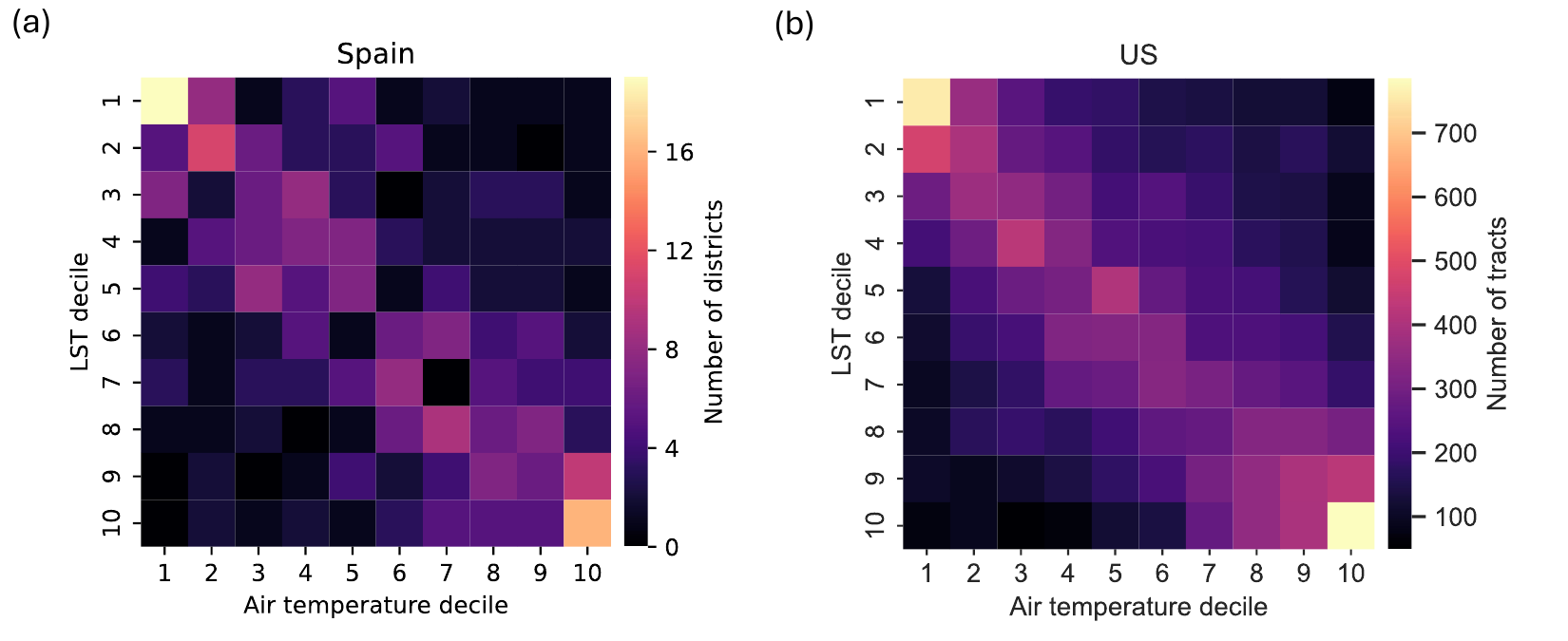} 
        \caption{\footnotesize \textbf{Comparison between land surface temperature (LST) and air temperature deciles.}  
        Comparison across all cities of district-level temperature deciles computed from mean-summer LST compared with deciles computed from mean-summer air temperature for (a) Spain and (b) US.}
\label{fig:SI_LST_vs_air_deciles_Spain_US}
\end{figure}

\begin{figure}[H]   
    \centering
    \includegraphics[width=\linewidth]{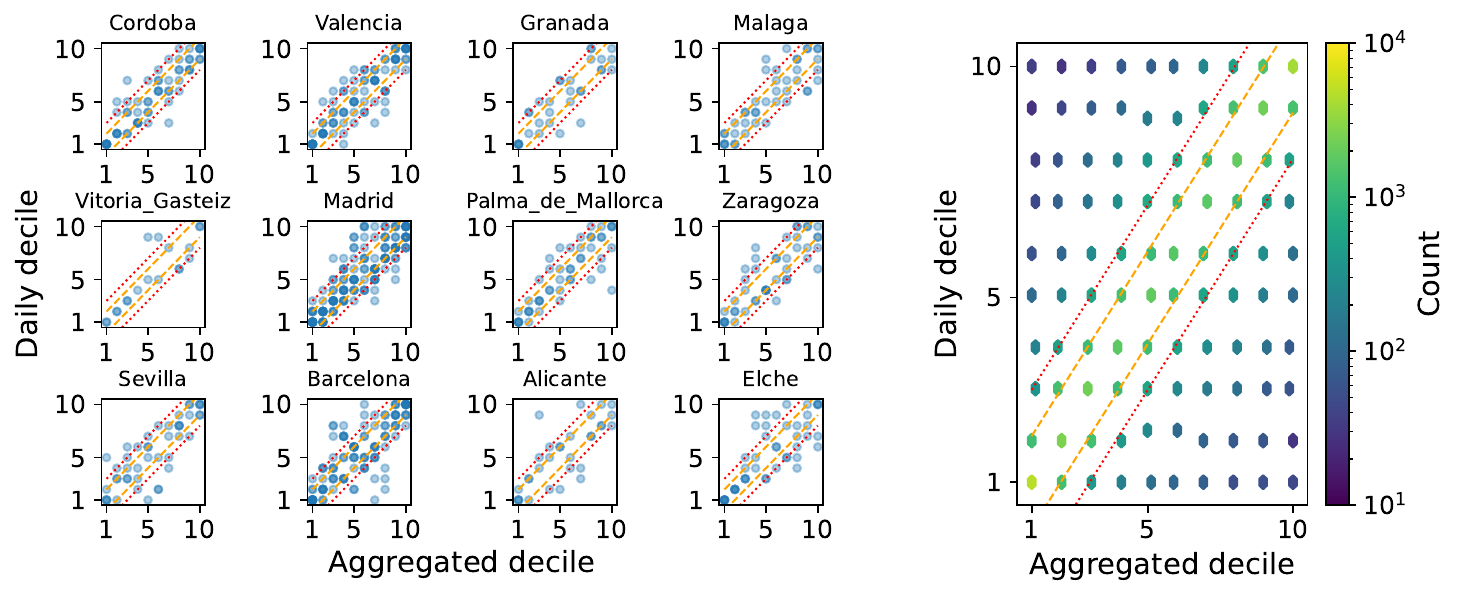} 
        \caption{\footnotesize \textbf{Comparison between aggregated and daily temperature deciles.}  
        On the left, for each city, district-level temperature deciles computed from mean-summer LST are compared with deciles computed for a single day. Each point represents a district’s aggregated decile versus its daily decile; the solid diagonal indicates perfect agreement, and dashed lines mark $\pm 1$ and $\pm 2$ deciles. 
        On the right pooled comparison across all cities, showing the count of district–day pairs in each aggregated–daily decile combination.}
\label{fig:SI_monthly_vs_daily_deciles}
\end{figure}

\begin{figure}[H]   
    \centering
    \includegraphics[width=0.8\linewidth]{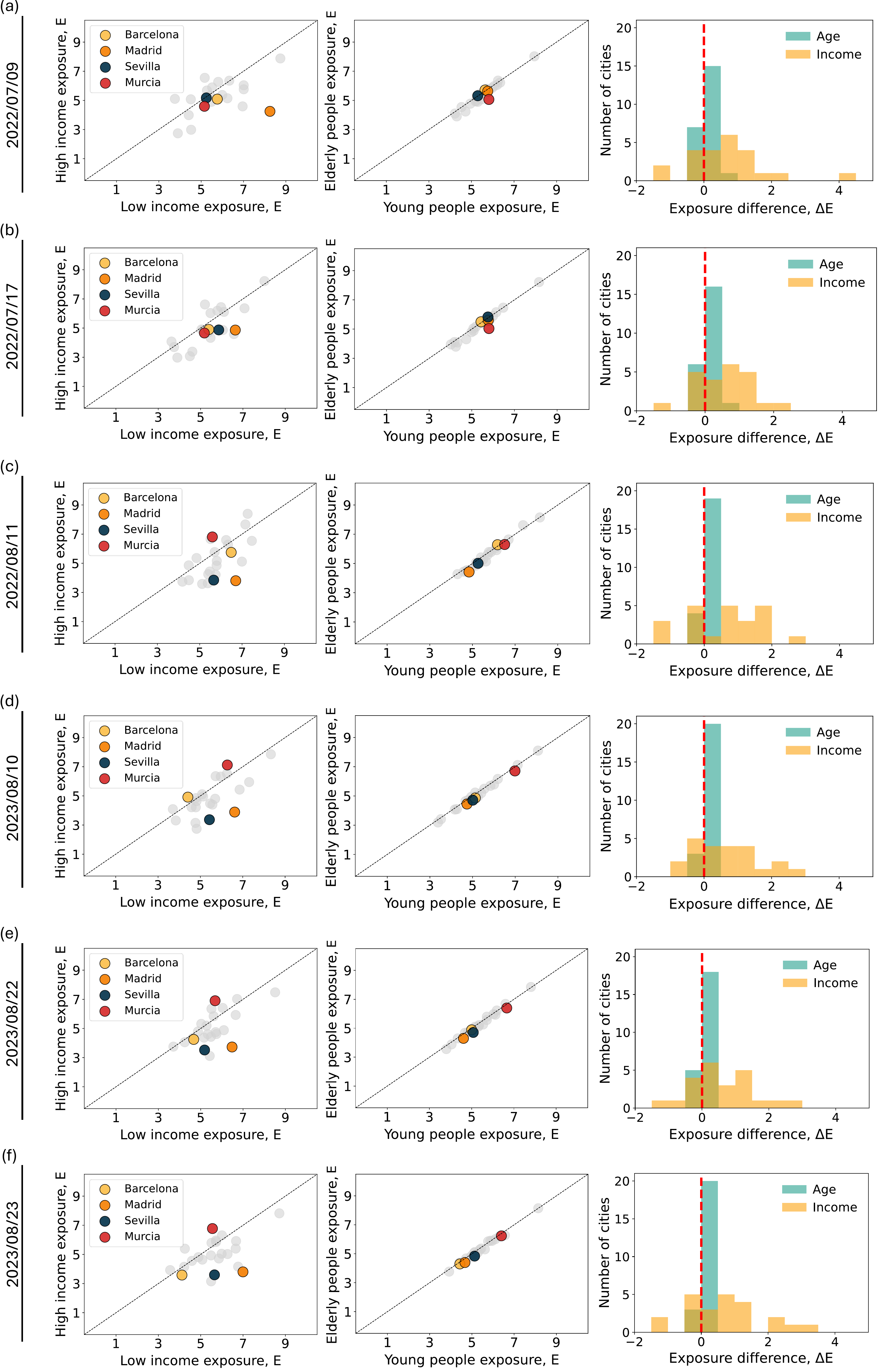} 
    \caption{\footnotesize 
    \textbf{Robustness test comparing aggregated versus daily temperature deciles for the mobility-based heat exposure inequality.} 
    To assess whether temporal aggregation biases the results, we recompute the exposure inequality for six randomly sampled days using daily temperature deciles: 
    (a) 2022/07/09, (b) 2022/07/17, (c) 2022/08/11, (d) 2023/08/10, (e) 2023/08/22, and (f) 2023/08/23. 
    Results closely match those obtained with aggregated summer deciles, confirming the stability of the exposure difference across sociodemographic groups and showing that discrepancies introduced by temporal aggregation are small and not systematically associated with any particular group.}

\label{fig:SI_random_days_test}
\end{figure}

\begin{figure}[H]   
    \centering
    \includegraphics[width=\linewidth]{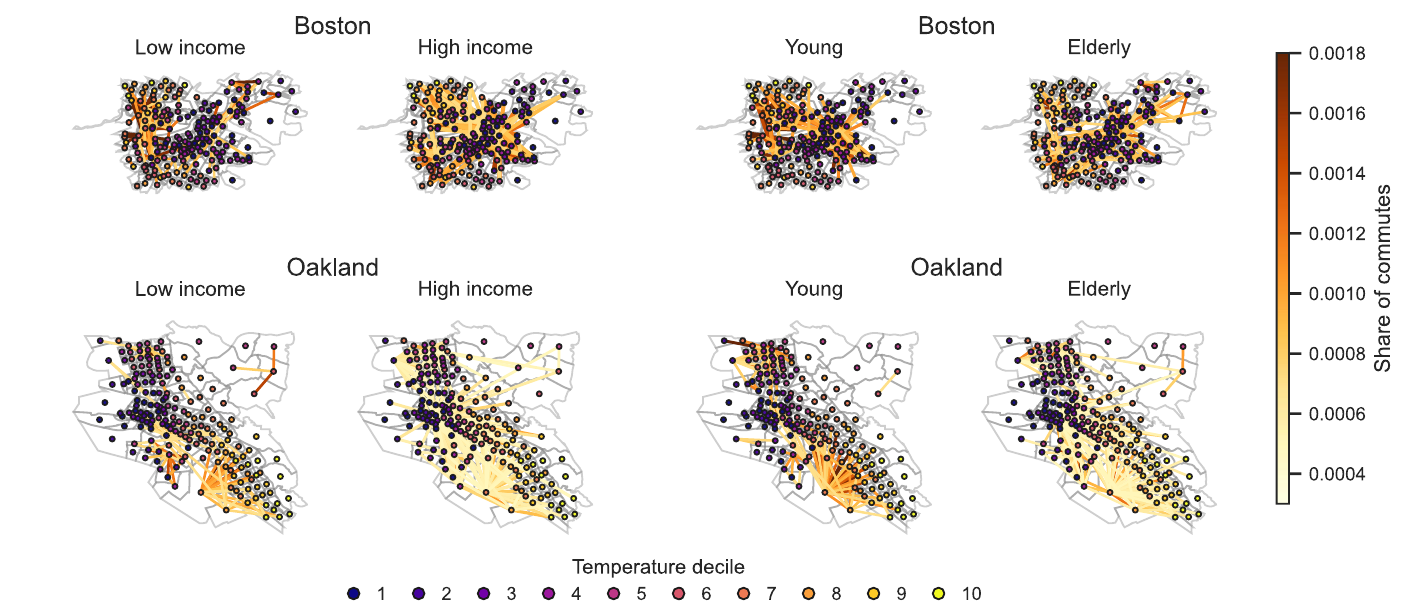} 
    \caption{\footnotesize
    \textbf{Illustrative home--work commuting networks stratified by income and age groups.}
    Tract-level home--work commuting networks for Boston and Oakland, constructed from 2022 LODES origin--destination flows. For each city, networks are shown separately for low-income groups and high-income groups, and young groups and elderly groups. Only the top $1.5\%$ of links by share of commuters are displayed, with edge color indicating the relative share of commuters. Nodes are colored by deciles of air temperature averaged over July--August 2022 and aggregated to Census tracts. These examples illustrate the commuting networks used to estimate heat exposure inequalities in the US external validation analysis. For visual clarity, panels show zoomed city windows.
    }
\label{fig:SI_illustrative_US}
\end{figure}

\begin{figure}[H]   
    \centering
    \includegraphics[width=\linewidth]{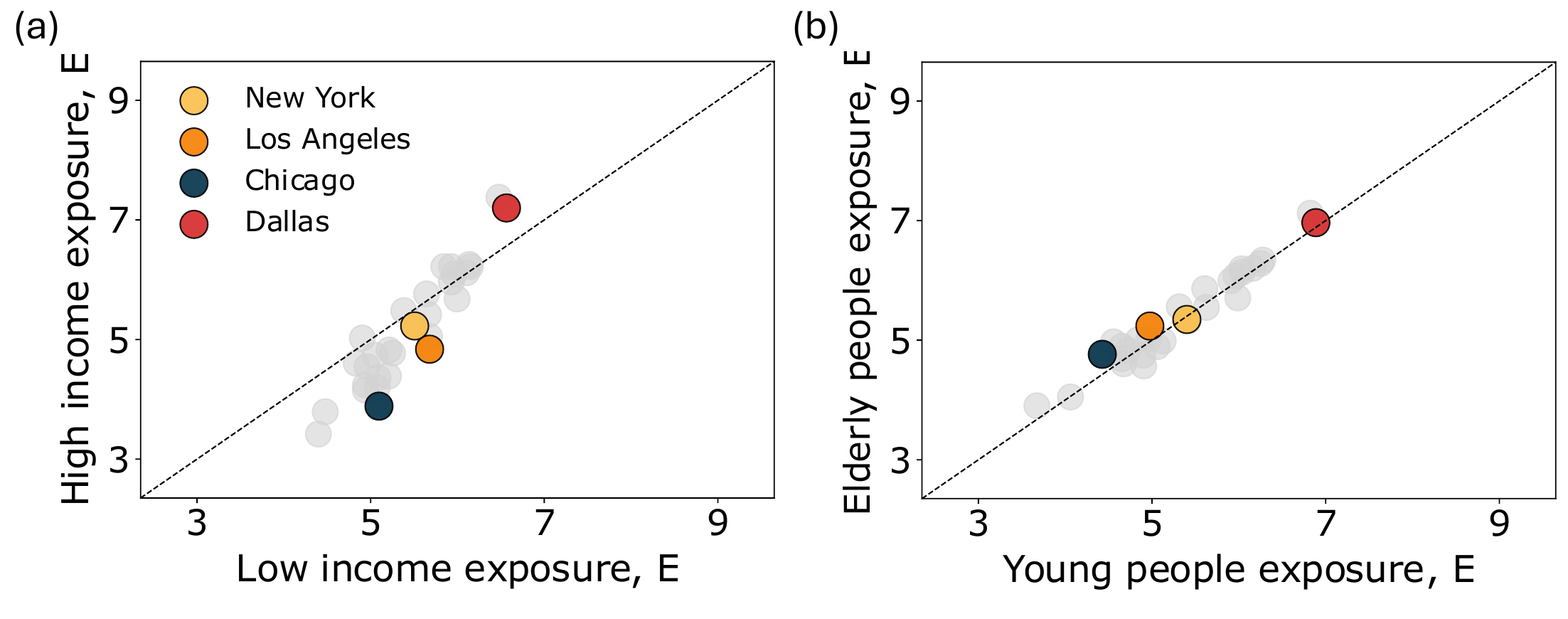} 
    \caption{\footnotesize \textbf{Inequalities in mobility-based heat exposure across income and age groups in US.}
    (a) Experienced heat exposure of high versus low-income groups, and (b) elderly versus young groups across cities. Colored points highlight selected cities for visual reference; grey points correspond to the remaining cities. The $y=x$ line indicates no exposure inequality between groups.}
\label{fig:SI_figure2_US}
\end{figure}

\begin{figure}[H]   
    \centering
    \includegraphics[width=\linewidth]{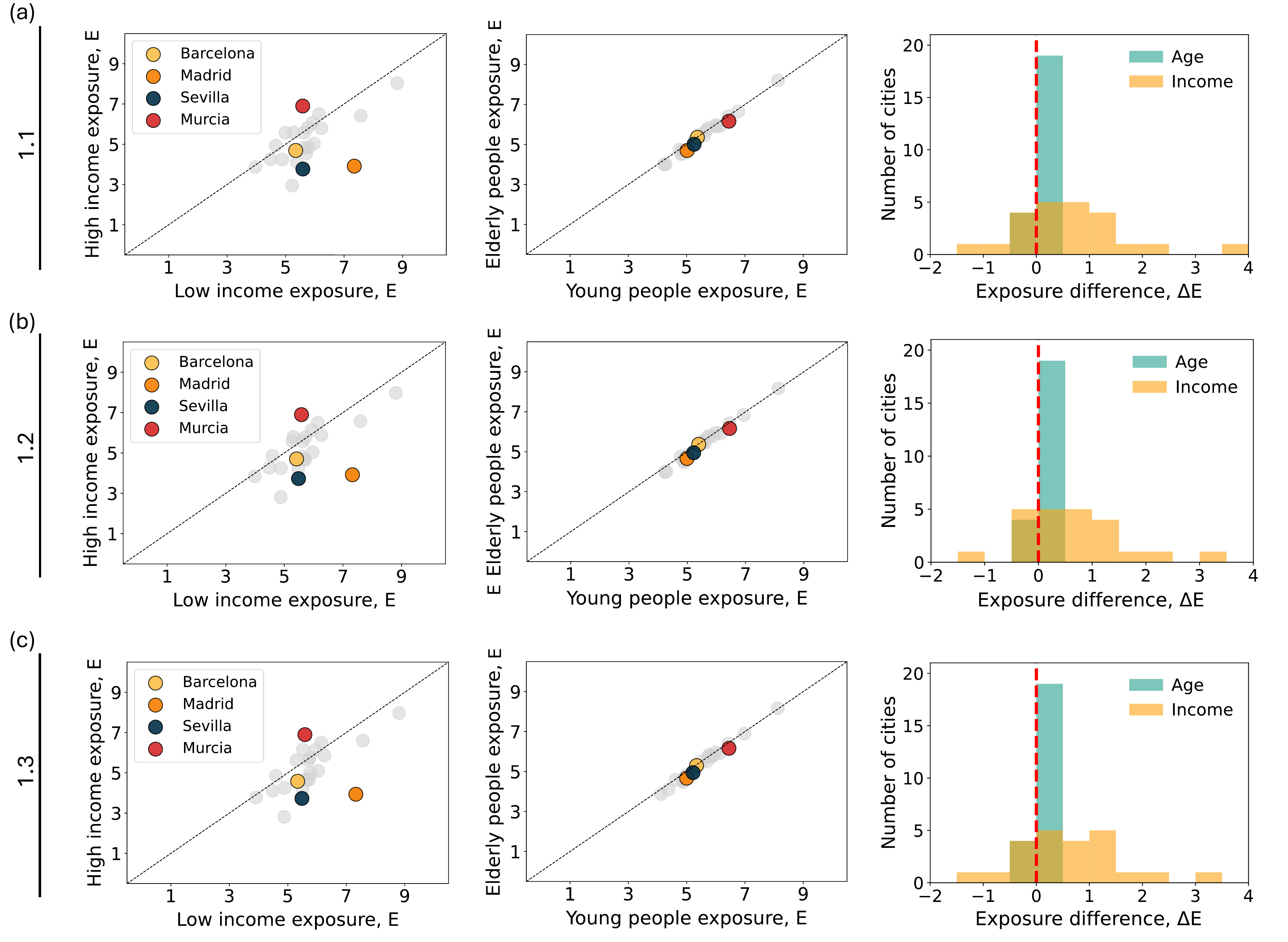} 
    \caption{\footnotesize 
    \textbf{Robustness of the inequality in mobility-based heat exposure to alternative definitions of city boundaries.} We compare results obtained using OpenStreetMap administrative boundaries with buffers of (a) 1.1× (b) 1.2× and (c) 1.3× the original extent. Across all buffer choices, the magnitude and sign of the income–heat exposure differences remain stable, indicating that the inequality signal is not sensitive to reasonable variations in the spatial definition of the urban area.}
\label{fig:SI_boundary_test}
\end{figure}

\begin{figure}[H]   
    \centering
    \includegraphics[width=\linewidth]{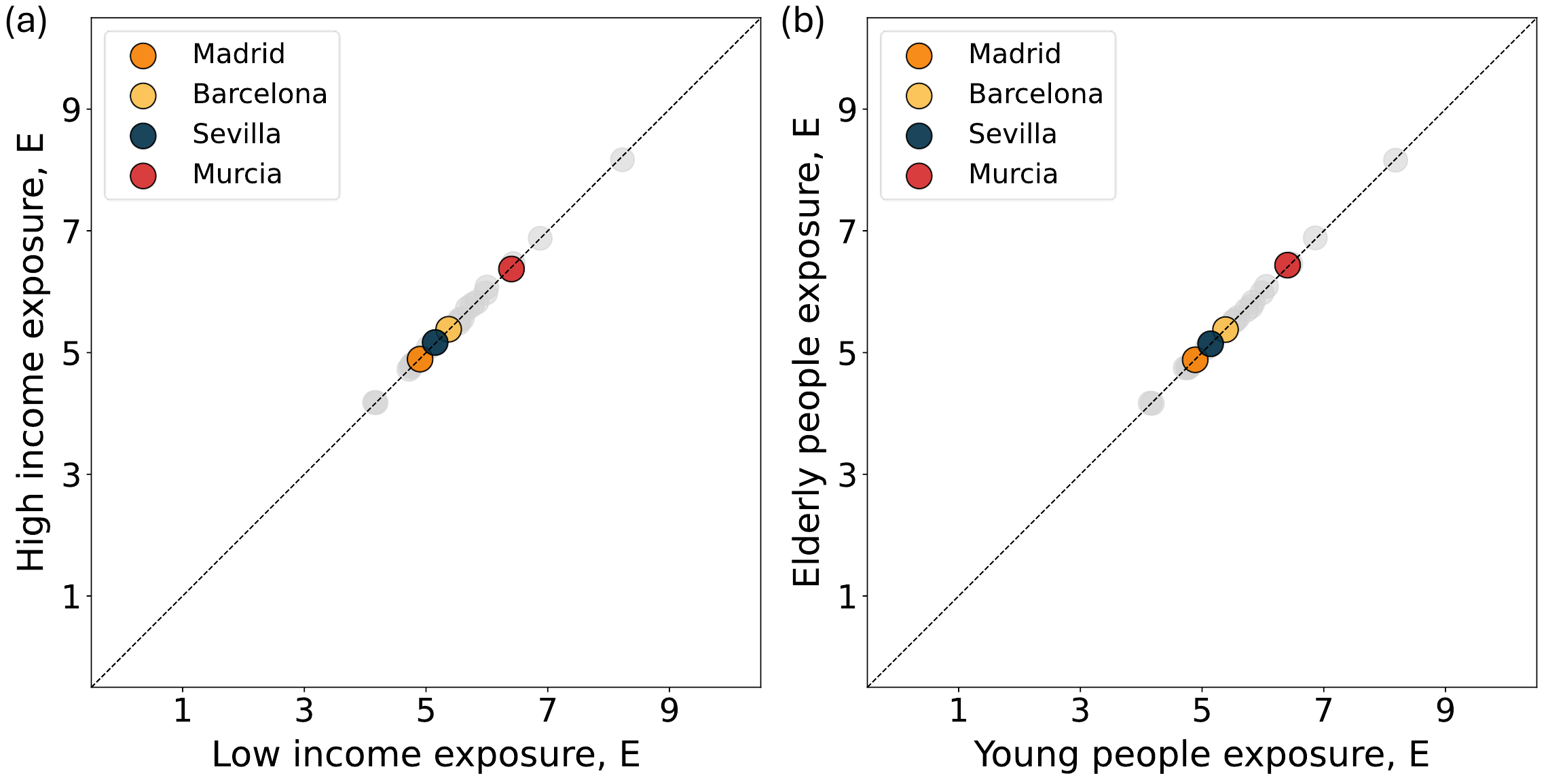} 
    \caption{\footnotesize 
    \textbf{Permutation test assessing whether inequality in mobility-based heat exposure could arise spuriously.} 
    We randomly shuffle (a) income and (b) age labels within each city–day while preserving the observed OD mobility structure and trip volumes. For each randomized realization ($B=100$ permutations), we recompute exposure differences between low- and high-income groups and between young and elderly populations. Scatter plots show one representative permutation. In all cases, randomized labels produce exposure gaps tightly centered around zero, whereas the observed inequalities lie far outside the null distributions (empirical permutation test: $p=0.01$). This indicates that the measured inequalities are unlikely to emerge from random assignment of sociodemographic labels within the mobility network.}
\label{fig:SI_permutation_test}
\end{figure}

\begin{figure}[H]   
    \centering
    \includegraphics[width=\linewidth]{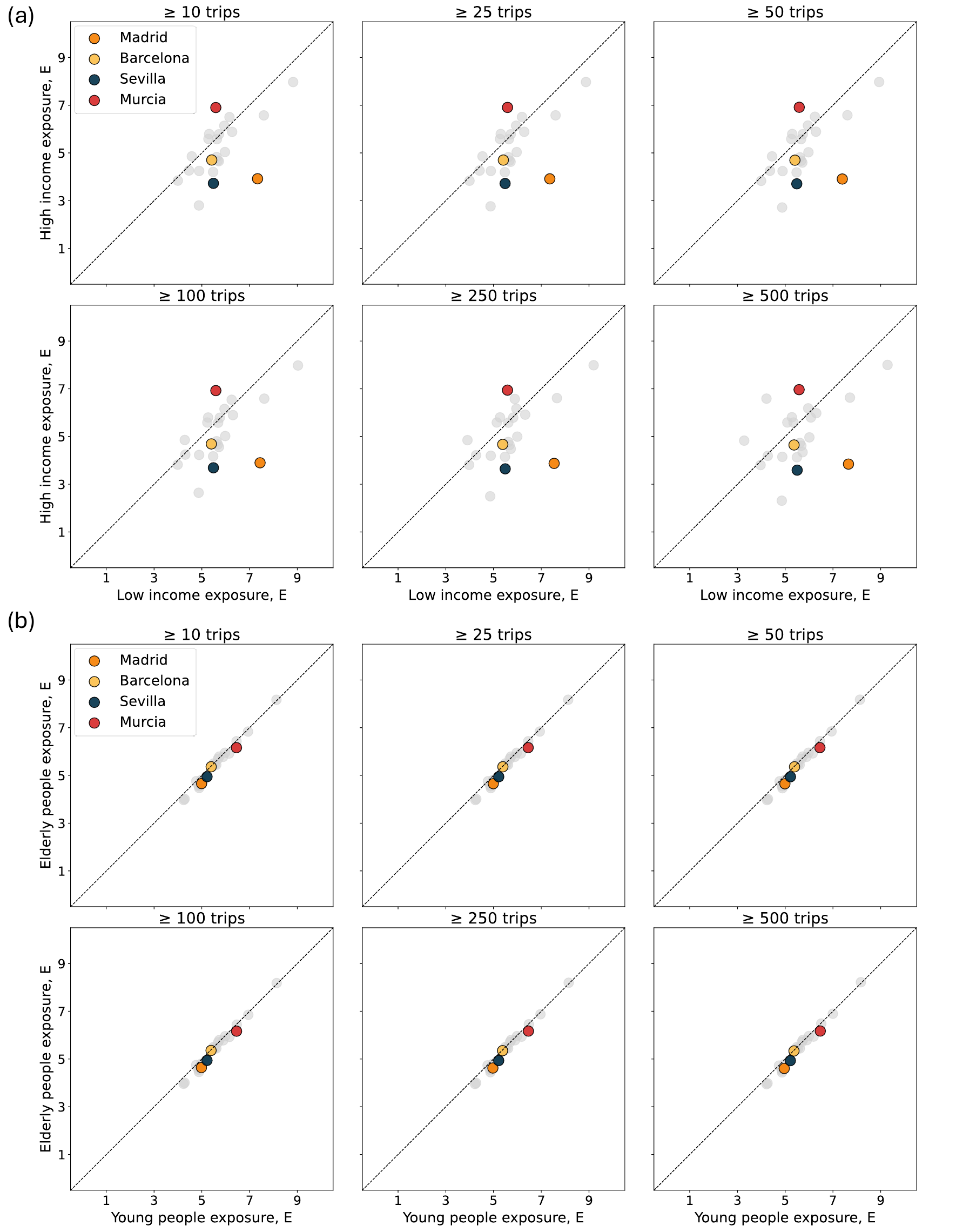} 
    \caption{\footnotesize 
    \textbf{Robustness of the inequality in mobility-based heat exposure to minimum trip thresholds in the mobility network.} 
    We recompute the heat exposure differences after applying increasingly strict minimum trip thresholds to remove low-volume, potentially noisy OD flows, for (a) income and (b) age groups. The magnitude and sign of the exposure differences remain stable across all threshold choices, indicating that the results are not sensitive to the treatment of low-volume mobility links.}
\label{fig:SI_thresholding_test}
\end{figure}


\begin{figure}[H]   
    \centering
    \includegraphics[width=\linewidth]{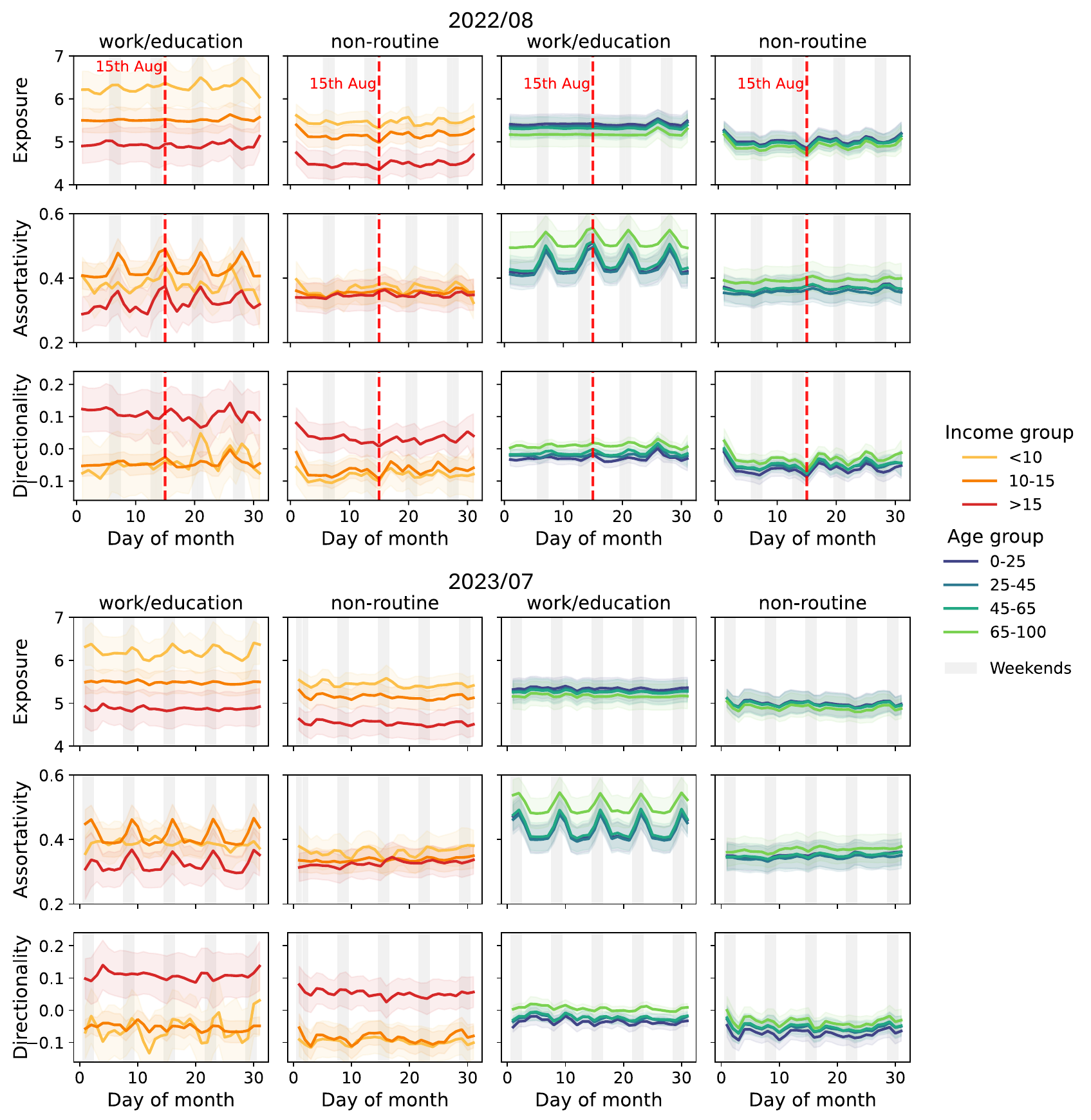} 
    \caption{\footnotesize 
    \textbf{Daily dynamics of mobility-based heat exposure, heat assortativity, and heat directionality differ between commuting and non-routine trips across income and age groups.}
    Daily population-weighted average of heat metrics across all cities in August 2022 (top panel) and in July 2022 (bottom panel). Error bands represent the standard error of the population-weighted average. Grey shading indicates weekends. The dashed red line marks 15 August. A 2-day centered window smooths short-term noise. }
\label{fig:SI_commute_vs_leisure}
\end{figure}

\begin{figure}[H]   
    \centering
    \includegraphics[width=\linewidth]{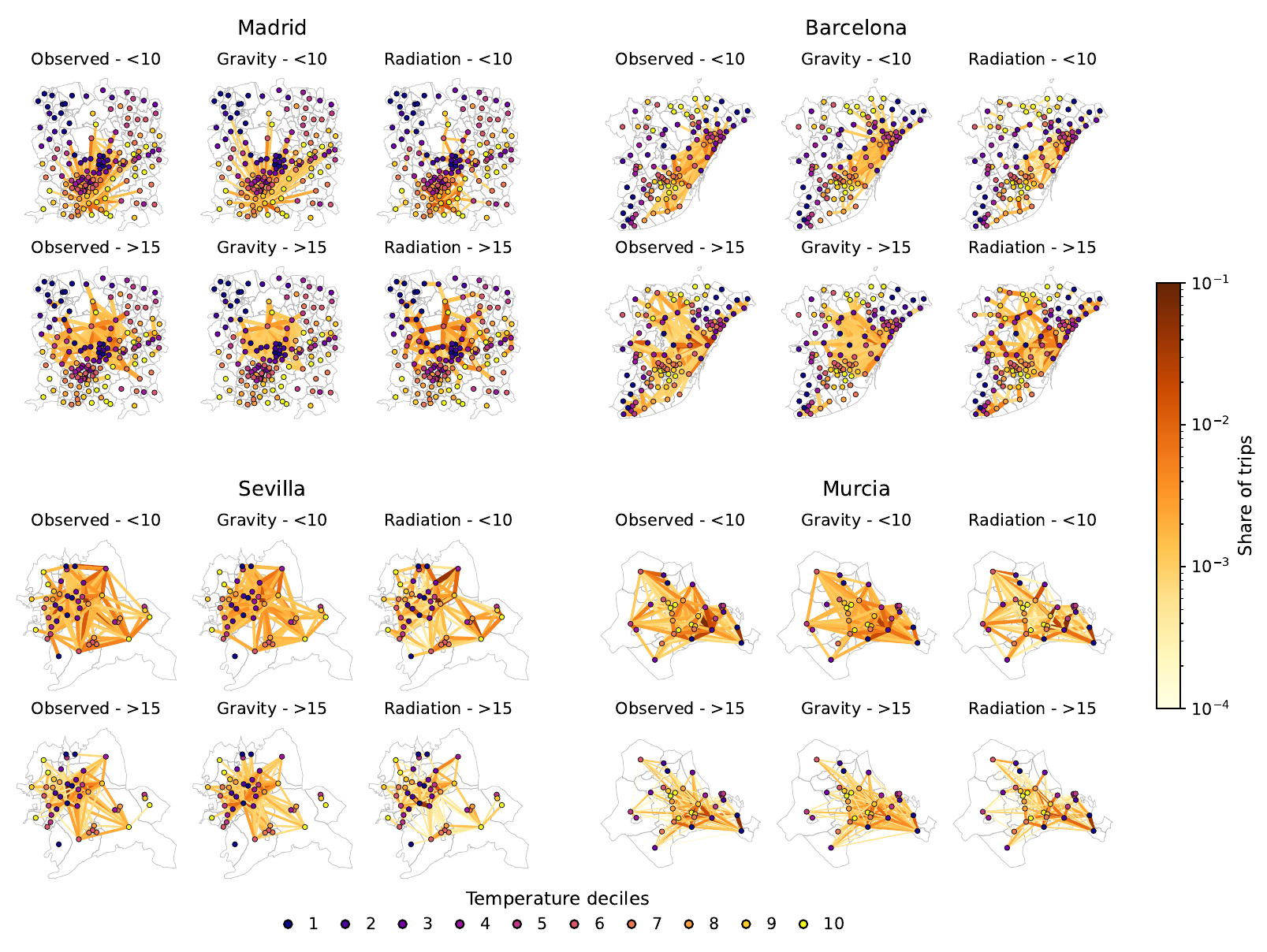} 
    \caption{\footnotesize 
    \textbf{Mobility networks by income group from observed and modeled origin--destination flows.} 
    Mobility networks derived from empirical data and from gravity and radiation models for four cities during July--August 2022 and July--August 2023. For each city and income group, only the top $200$ links ranked by share of trips are displayed. Edge width scales with the absolute number of trips, while edge color encodes the relative share of trips. Nodes are colored by deciles of average district temperature over the same period.}
\label{fig:SI_flows_obs_modeled_income}
\end{figure}

\begin{figure}[H]   
    \centering
    \includegraphics[width=\linewidth]{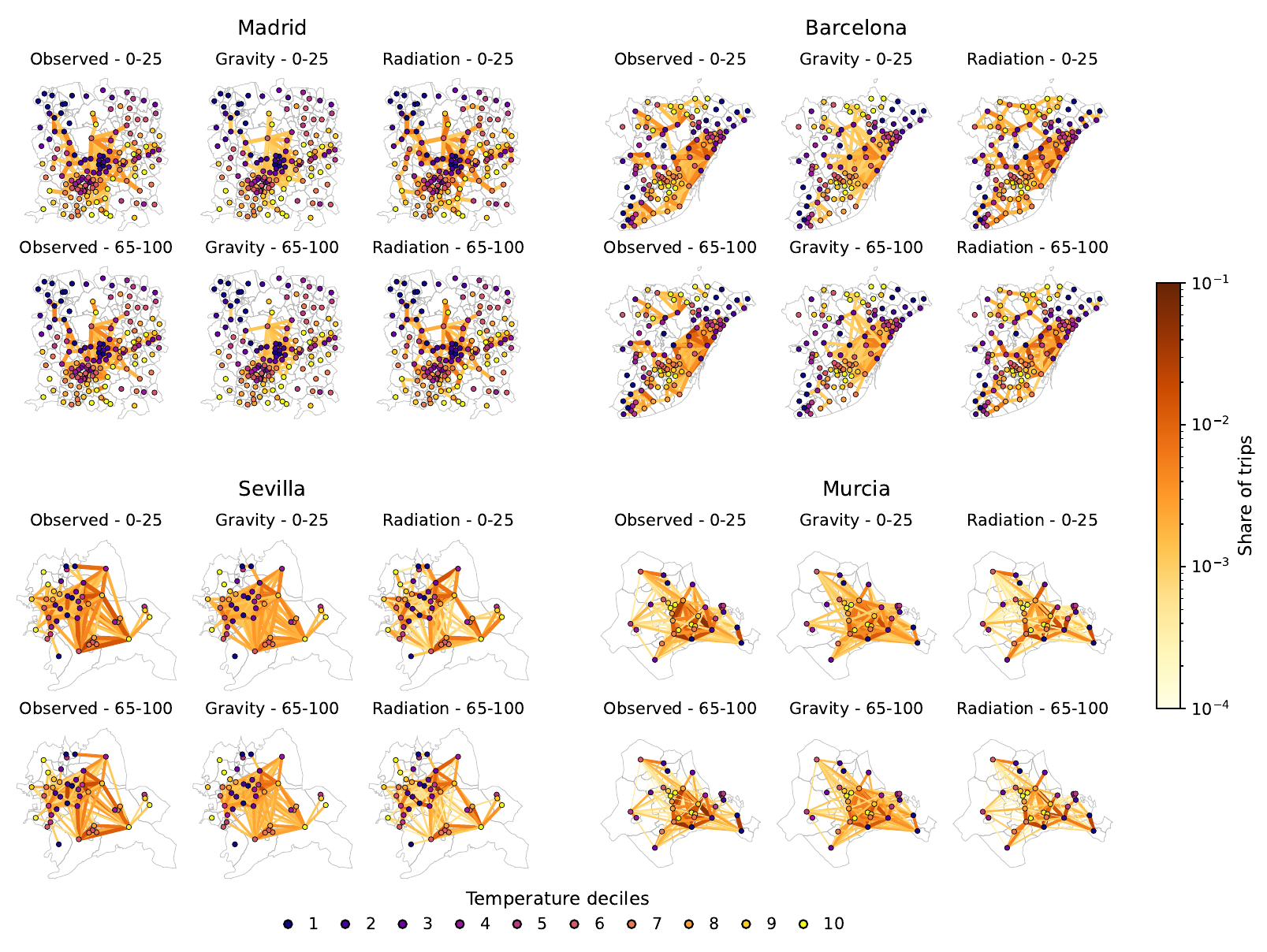} 
    \caption{\footnotesize 
    \textbf{Mobility networks by age group from observed and modeled origin--destination flows.} 
    Mobility networks derived from empirical data and from gravity and radiation models for four cities during July--August 2022 and July--August 2023. For each city and income group, only the top $200$ links ranked by share of trips are displayed. Edge width scales with the absolute number of trips, while edge color encodes the relative share of trips. Nodes are colored by deciles of average district temperature over the same period.}
\label{fig:SI_flows_obs_modeled_age}
\end{figure}


\begin{figure}[H]   
    \centering
    \includegraphics[width=\linewidth]{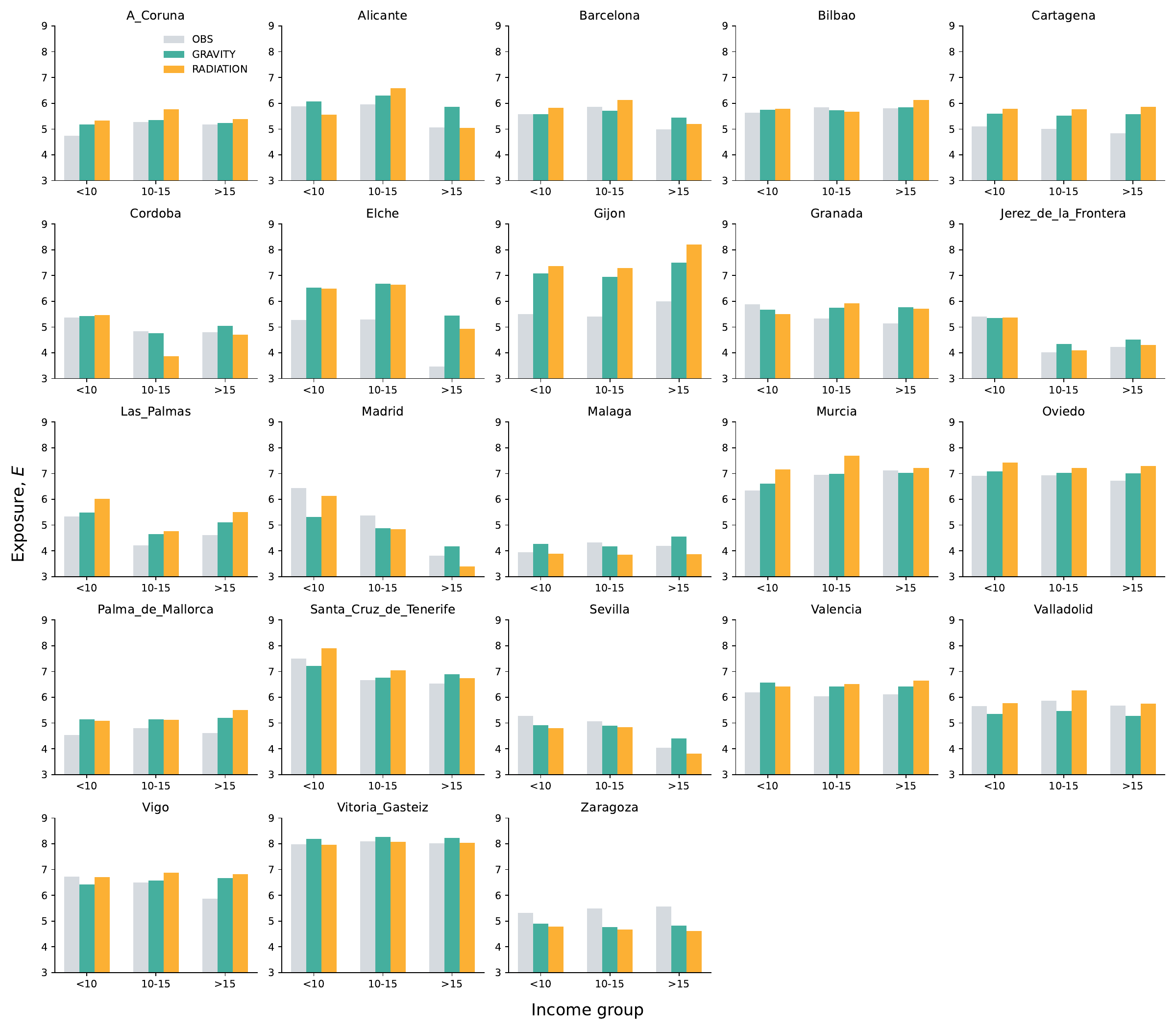} 
    \caption{\footnotesize 
    \textbf{Exposure $E$ of observed and modeled mobility networks by income group per city.}
    The $E$ values are computed from district-level origin--destination networks for observed data (grey) and for gravity (green) and radiation (orange) models. For comparability with the models, both observed and modeled assortativity values are computed after excluding intra-district flows.}

\label{fig:SI_E_income_per_city}
\end{figure}

\begin{figure}[H]   
    \centering
    \includegraphics[width=\linewidth]{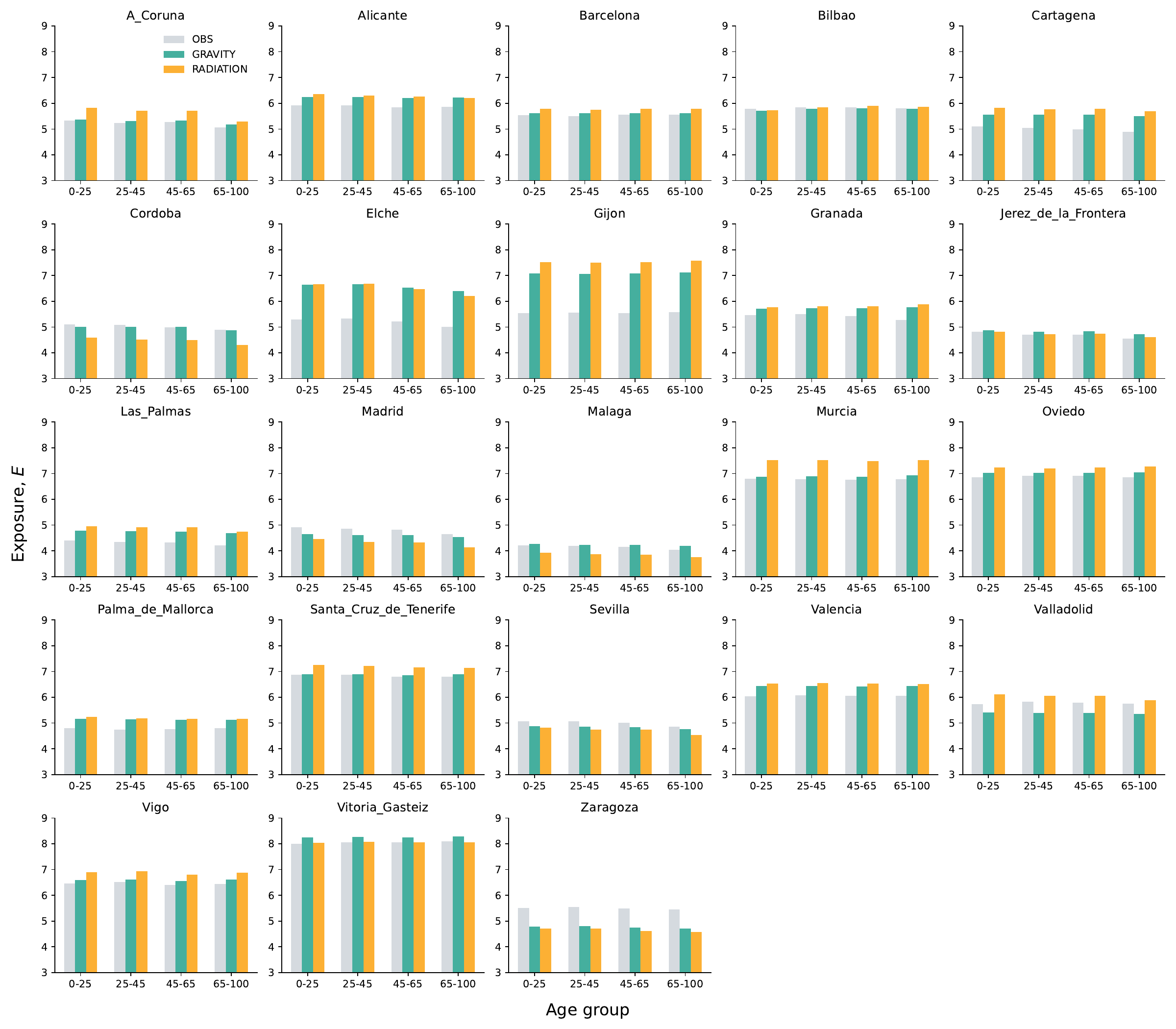} 
    \caption{\footnotesize 
    \textbf{Exposure $E$ of observed and modeled mobility networks by age group per city.}
    The $E$ values are computed from district-level origin--destination networks for observed data (grey) and for gravity (green) and radiation (orange) models. For comparability with the models, both observed and modeled assortativity values are computed after excluding intra-district flows.}

\label{fig:SI_E_age_per_city}
\end{figure}

\begin{figure}[H]   
    \centering
    \includegraphics[width=\linewidth]{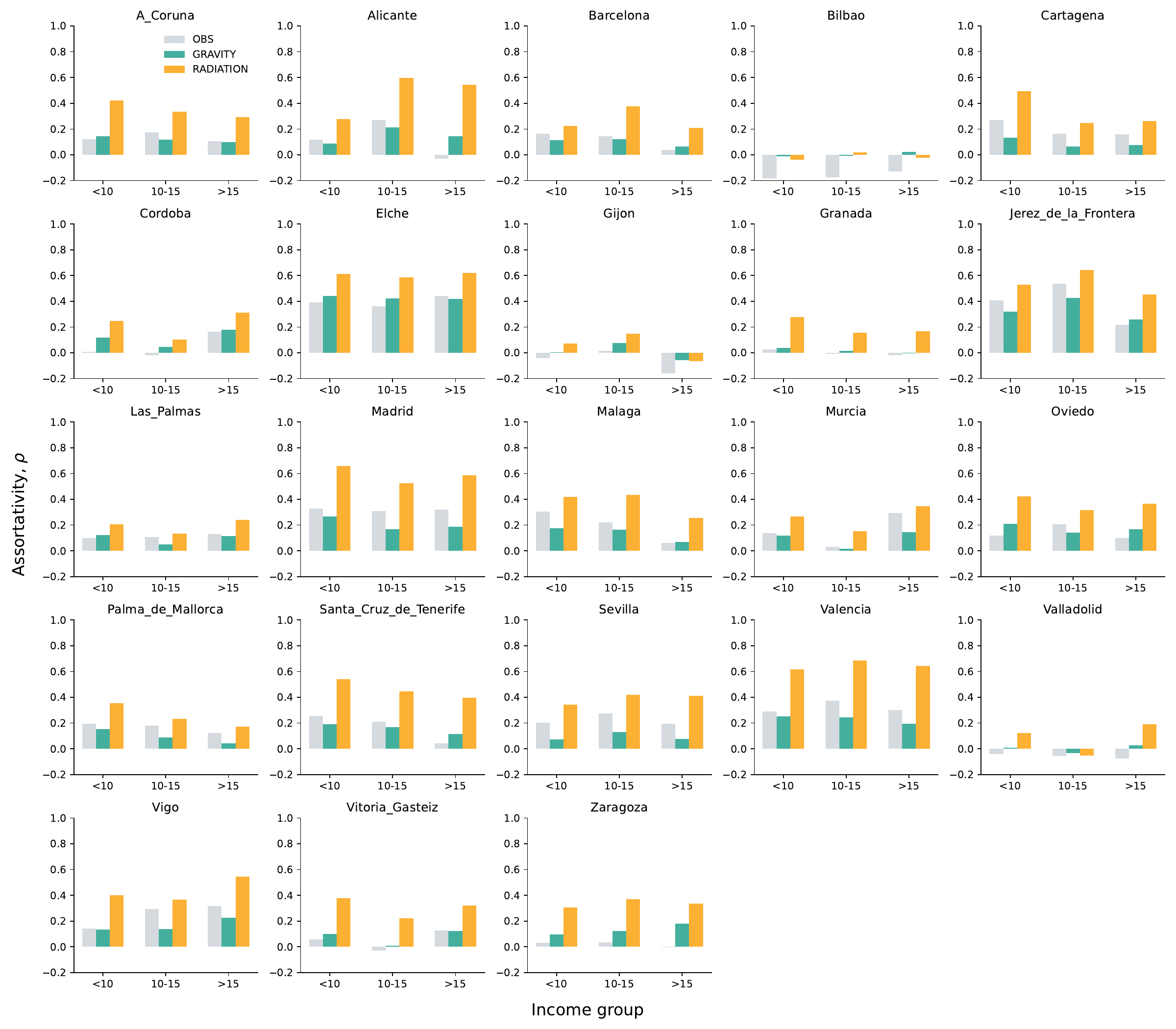} 
    \caption{\footnotesize 
    \textbf{Assortativity coefficient $\rho$ of observed and modeled mobility networks by income group per city.}
    The $\rho$ values are computed from district-level origin--destination networks for observed data (grey) and for gravity (green) and radiation (orange) models. For comparability with the models, both observed and modeled assortativity values are computed after excluding intra-district flows.}

\label{fig:SI_rho_income_per_city}
\end{figure}

\begin{figure}[H]   
    \centering
    \includegraphics[width=\linewidth]{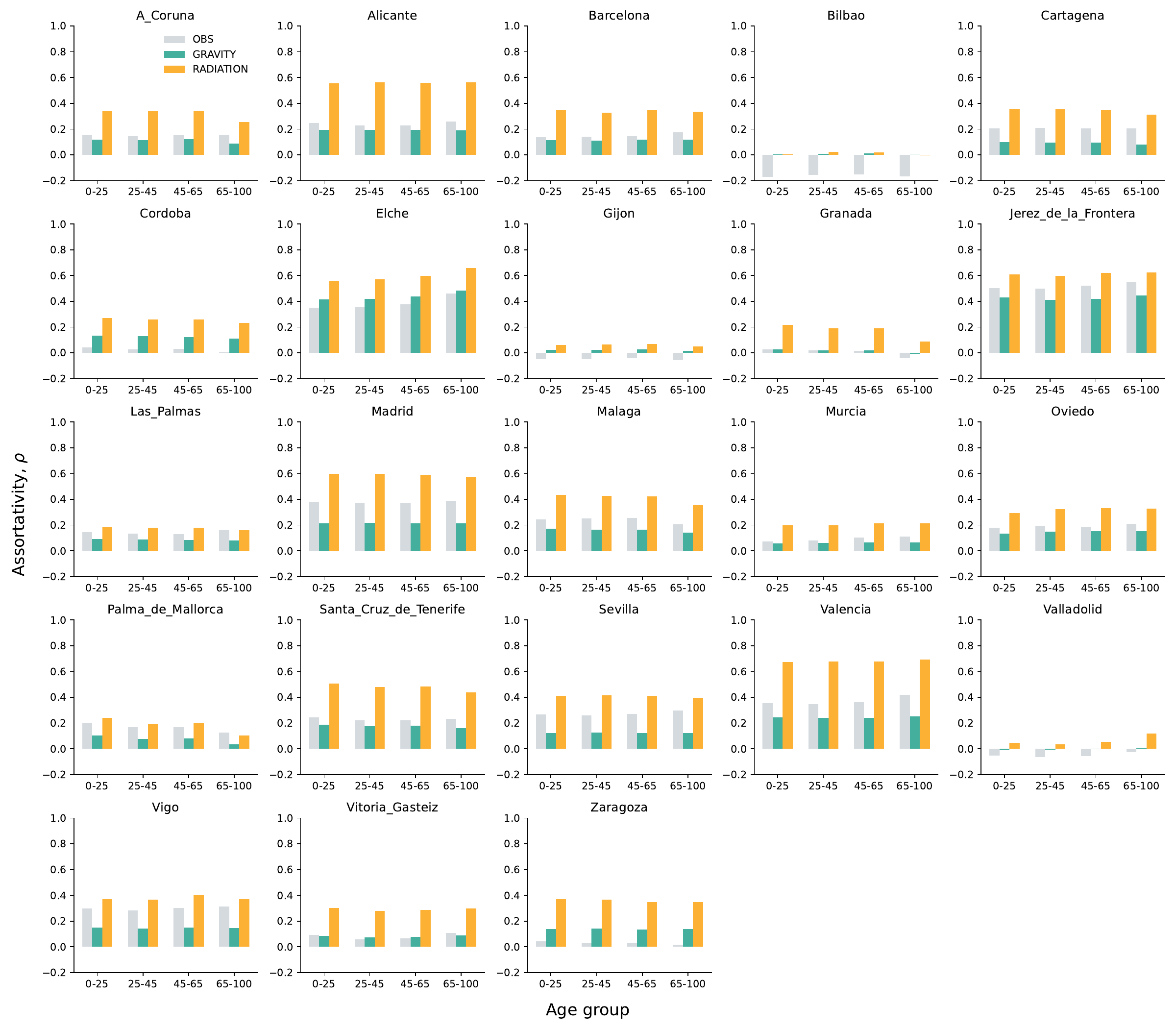} 
    \caption{\footnotesize 
    \textbf{Assortativity coefficient $\rho$ of observed and modeled mobility networks by age group per city.}
    The $\rho$ values are computed from district-level origin--destination networks for observed data (grey) and for gravity (green) and radiation (orange) models. For comparability with the models, both observed and modeled assortativity values are computed after excluding intra-district flows.}

\label{fig:SI_rho_age_per_city}
\end{figure}

\begin{figure}[H]   
    \centering
    \includegraphics[width=\linewidth]{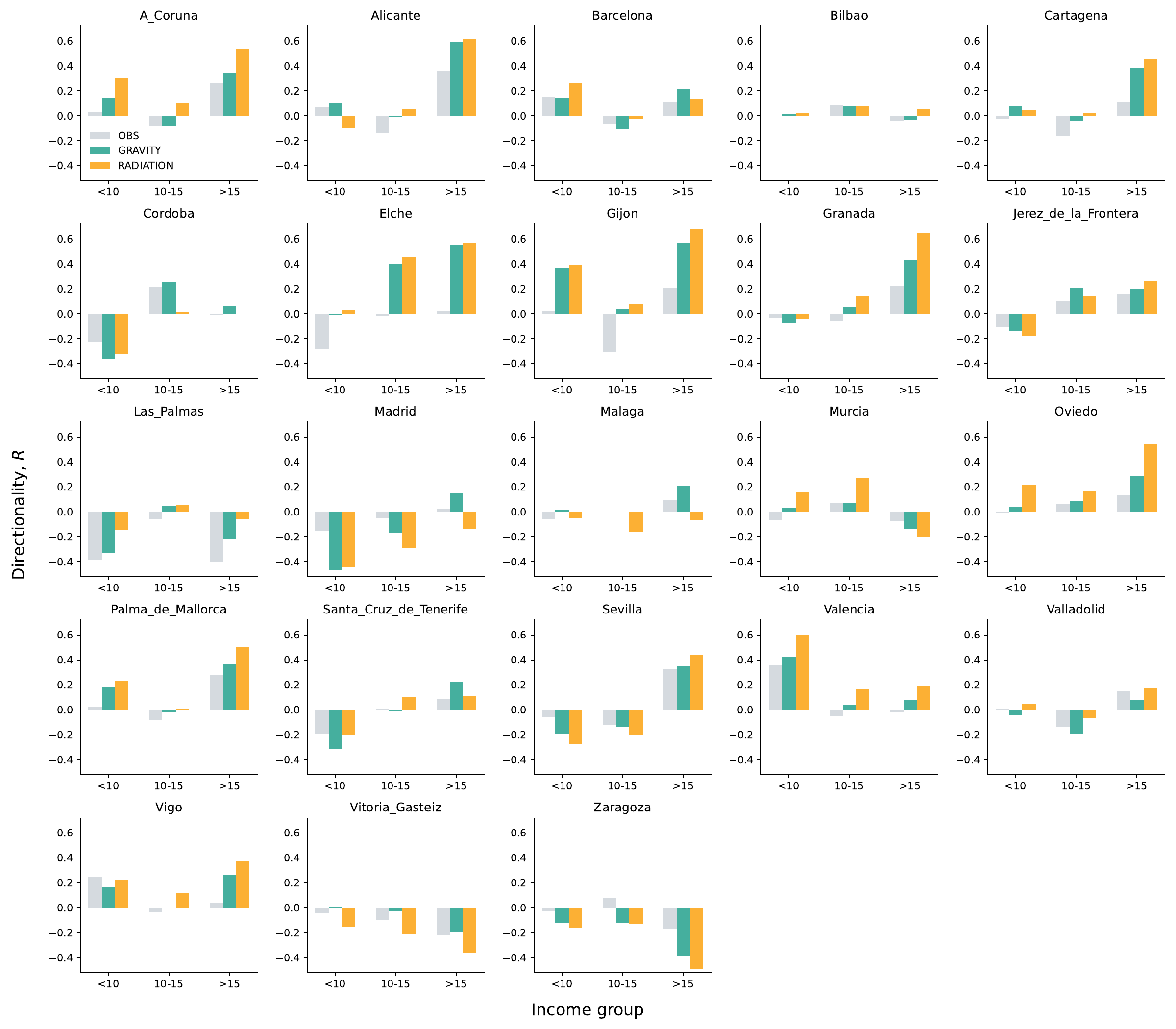} 
    \caption{\footnotesize 
    \textbf{Directionality index $R$ of observed and modeled mobility networks by income group per city.}
    The $R$ values are computed from district-level origin--destination networks for observed data (grey) and for gravity (green) and radiation (orange) models. For comparability with the models, both observed and modeled assortativity values are computed after excluding intra-district flows.}

\label{fig:SI_R_income_per_city}
\end{figure}

\begin{figure}[H]   
    \centering
    \includegraphics[width=\linewidth]{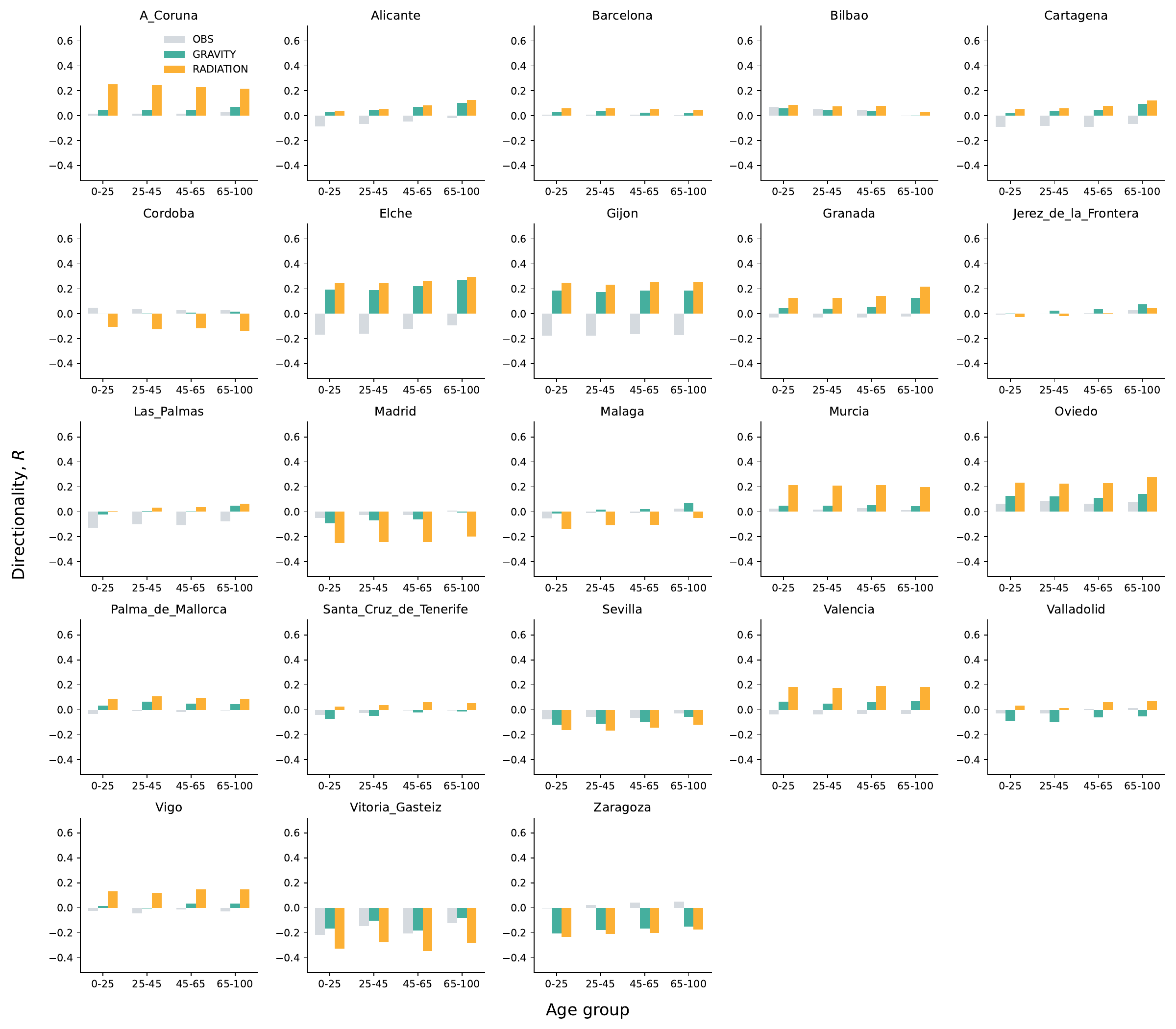} 
    \caption{\footnotesize 
    \textbf{Directionality index $R$ of observed and modeled mobility networks by age group per city.}
    The $R$ values are computed from district-level origin--destination networks for observed data (grey) and for gravity (green) and radiation (orange) models. For comparability with the models, both observed and modeled assortativity values are computed after excluding intra-district flows.}

\label{fig:SI_R_age_per_city}
\end{figure}



\begin{figure}[H]   
    \centering
    \includegraphics[width=\linewidth]{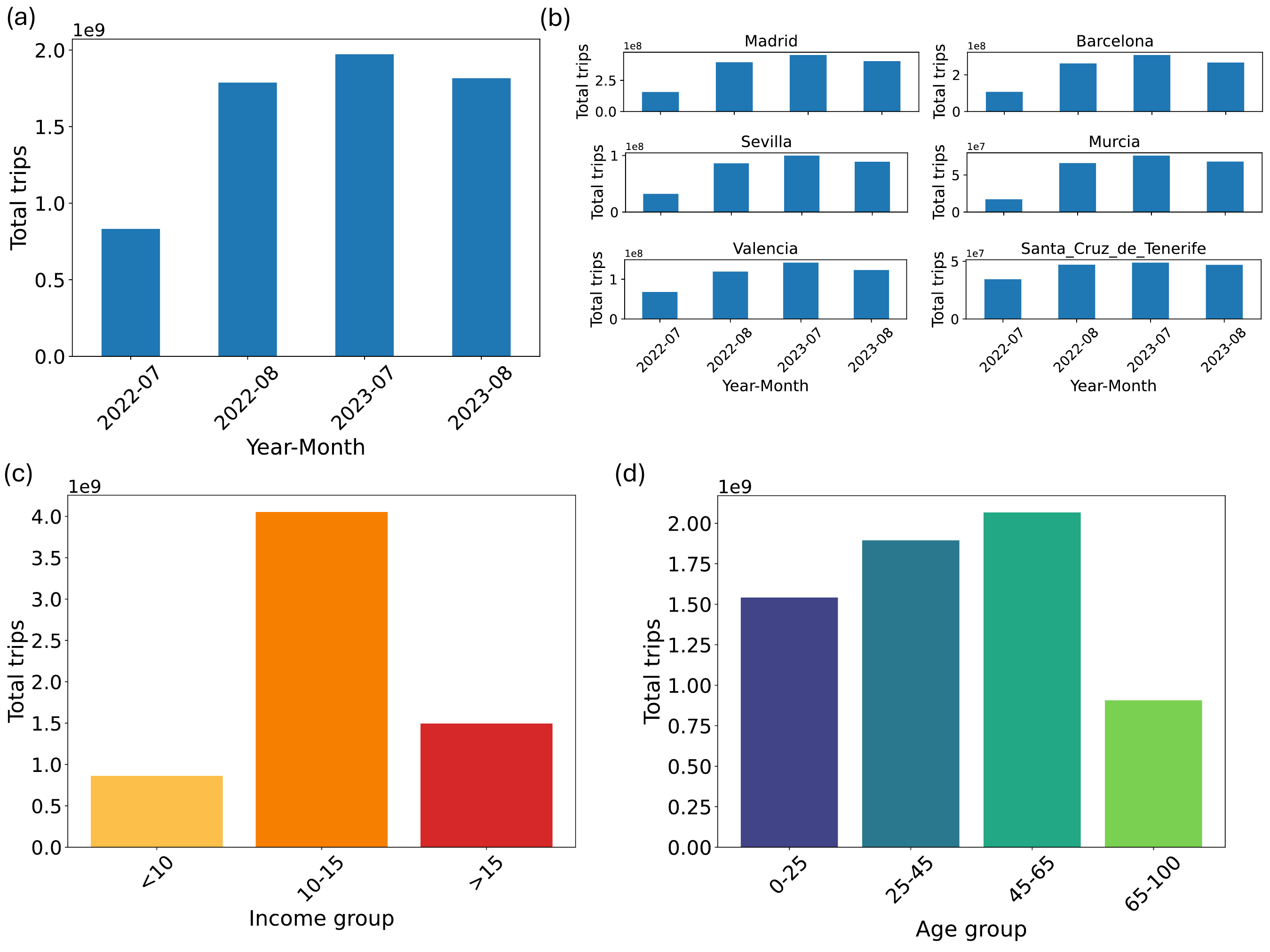} 
        \caption{\footnotesize \textbf{Total number of trips during July-August 2022 and 2023.}  (a) Total number of trips aggregated across all cities. (b) Total number of trips for a selected subset of cities. Total number of trips aggregated across all cities, stratified by (c) income and (d) age.}
\label{fig:SI_total_number_of_trips}
\end{figure}

\begin{figure}[H]   
    \centering
    \includegraphics[width=\linewidth]{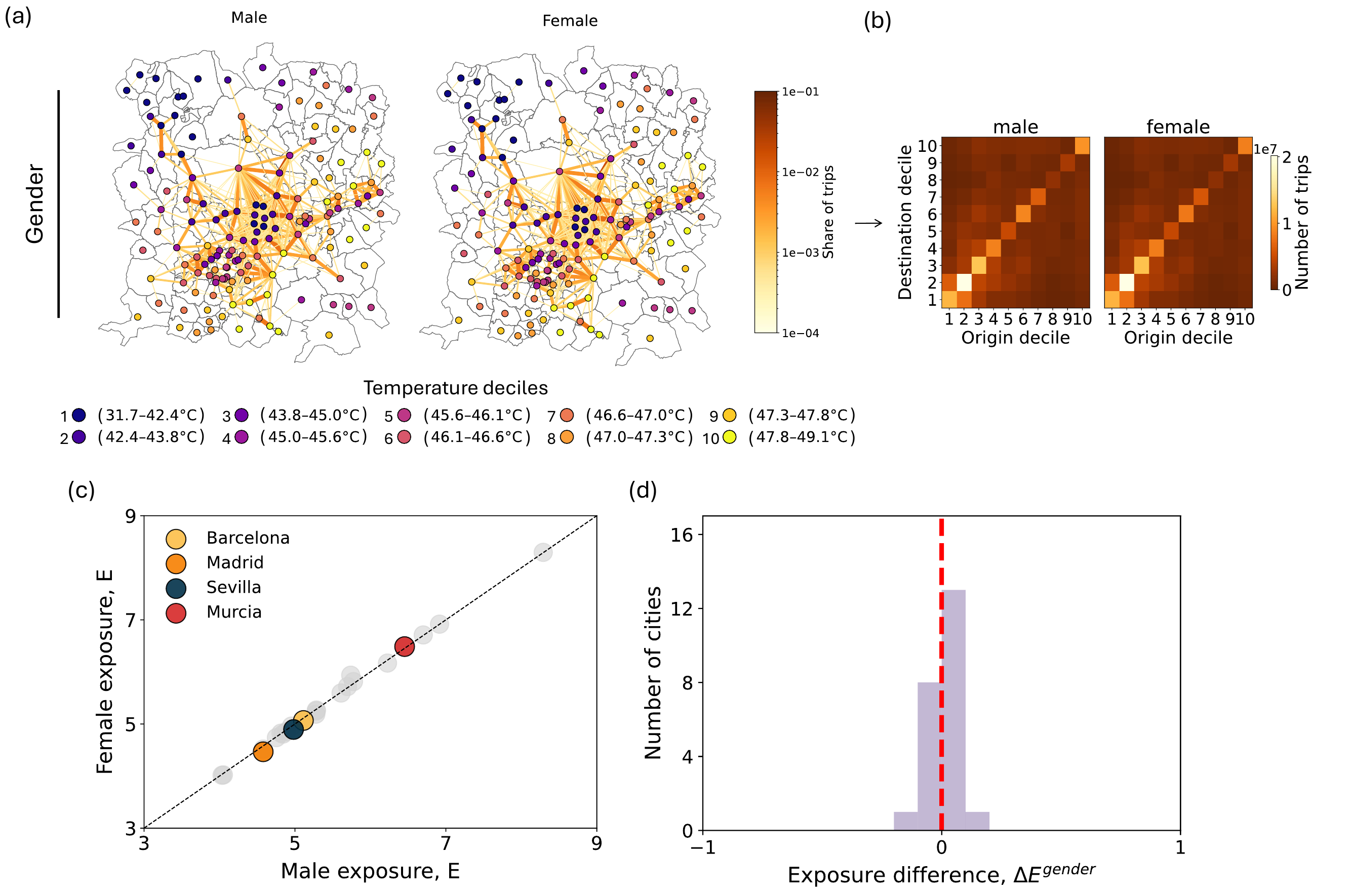} 
        \caption{\footnotesize \textbf{No detectable gender inequality in mobility-based heat exposure.}  
        (a) Mobility networks at the Spanish district level stratified by gender groups in Madrid, August 2023.
        (b) Origin–destination (OD) matrices between the ten temperature deciles for gender groups in Madrid. 
        (c) Scatter plot of average exposure for male versus female groups, where each point represents one city. 
        (d) Histogram of the exposure difference $\Delta E^{gender}$ for gender groups.}
\label{fig:SI_gender}
\end{figure}


\end{document}